\documentclass[oneside,a4paper,11pt,explicit]{book}
\def\ARXIVBUILD{1}
\ifdefined\ARXIVBUILD
\else
  \csname input\endcsname{arxiv_volume_IV.tex}
\fi

\usepackage[T1]{fontenc}
\usepackage[utf8]{inputenc}
\usepackage{textcomp}
\usepackage{vol4_kultem}

\usepackage{amsmath,amssymb}
\usepackage{booktabs}
\usepackage{tabularx}
\usepackage{array}
\usepackage{longtable}
\usepackage{lscape}
\usepackage{graphicx}
\usepackage[dvipsnames,svgnames,table]{xcolor}
\usepackage[most]{tcolorbox}
\usepackage{float}
\usepackage{hyperref}

\makeatletter
\renewcommand*\l@section{\@dottedtocline{1}{1.5em}{2.8em}}
\renewcommand*\l@subsection{\@dottedtocline{2}{3.8em}{3.7em}}
\renewcommand*\l@subsubsection{\@dottedtocline{3}{7.0em}{4.6em}}
\makeatother

\usetikzlibrary{arrows.meta,calc,positioning,shapes.geometric,
  decorations.pathmorphing,backgrounds,fit}
\numberwithin{equation}{chapter}

\definecolor{ink}{HTML}{151B23}
\definecolor{titlebg}{HTML}{100880}
\definecolor{softink}{HTML}{263441}
\definecolor{muted}{HTML}{5C6875}
\definecolor{paper}{HTML}{F6F7F2}
\definecolor{panel}{HTML}{FFFFFF}
\definecolor{line}{HTML}{D8DEE5}
\definecolor{teal}{HTML}{007C77}
\definecolor{blue}{HTML}{245AA6}
\definecolor{gold}{HTML}{B86B00}
\definecolor{rose}{HTML}{A7354D}
\definecolor{green}{HTML}{23724A}
\definecolor{violet}{HTML}{5A4CA0}

\newcommand{\facility}{3.5-meter Segmented-Mirror Robotic Space Telescope}

\newcolumntype{Y}{>{\raggedright\arraybackslash}X}

\newtcolorbox{leadbox}[2][]{
  enhanced, colback=paper, colframe=#2, boxrule=0.9pt, arc=2mm,
  left=2.2mm, right=2.2mm, top=1.8mm, bottom=1.8mm,
  fonttitle=\sffamily\bfseries, coltitle=white,
  attach boxed title to top left={xshift=2mm,yshift=-2mm},
  boxed title style={colback=#2,arc=1.2mm,boxrule=0pt}, #1 }
\newtcolorbox{metricbox}[1]{
  enhanced, colback=white, colframe=#1, boxrule=0.65pt, arc=1.3mm,
  left=1.8mm, right=1.8mm, top=1.6mm, bottom=1.6mm }

\makeatletter
\renewenvironment{thebibliography}[1]
  {\section*{References}\@mkboth{}{}%
   \list{\@biblabel{\@arabic\c@enumiv}}%
        {\settowidth\labelwidth{\@biblabel{#1}}%
         \leftmargin\labelwidth \advance\leftmargin\labelsep
         \usecounter{enumiv}\let\p@enumiv\@empty
         \renewcommand\theenumiv{\@arabic\c@enumiv}}%
   \small\sloppy\clubpenalty4000\widowpenalty4000\sfcode`\.\@m}
  {\def\@noitemerr{\@latex@warning{Empty `thebibliography' environment}}\endlist}
\makeatother

\renewcommand{\topfraction}{0.9}

\renewcommand{\floatpagefraction}{0.85}
\newcommand{\wpvolumelabel}{IV}
\newcommand{\wpvolumetitle}{Key Scientific Mission: Solar-System Small Bodies and Planetary Defense}
\title{3.5-meter Segmented-Mirror Robotic Space Telescope}
\subtitle{Mission White Paper: \wpvolumelabel. \wpvolumetitle}
\newcommand{\wpauthors}{Juhan Kim$^{1}$, Yong-Woo Kang$^{2}$, Sang Hyun Lee$^{2,3}$, Jeong-Yeol Han$^{2,4}$, Sungwook E. Hong$^{2,4}$, Bongkon Moon$^{2,4}$, Donguk Song$^{2}$, Juhyung Kang$^{2}$, Myeong-Gu Park$^{5}$, Sang Chul Kim$^{2,4}$, Chung-Uk Lee$^{2}$, Sangmo Tony Sohn$^{6}$, Arman Shafieloo$^{2,4}$, David Parkinson$^{2,4}$, Hong Soo Park$^{2,4}$, Dohyeong Kim$^{7}$, Chan Park$^{2}$, Jungjoo Sohn$^{8}$, Young-Beom Jeon$^{2}$, Jong-Hak Woo$^{9}$, Hyung Mok Lee$^{9}$, Hong Bae Ann$^{7}$, Myungkook James Jee$^{10}$, Mansoo Choi$^{2}$, Changbom Park$^{1}$}
\date{2026}
\newcommand{\wpabstract}{%
The baseline 0.2--1.5\,$\mu$m observatory provides rapid-response astrometry, visible and near-infrared taxonomy, rotation and phase curves, recovery, and long-arc orbit improvement for near-Earth objects and other small bodies. The instrument study also evaluates calibrated throughput to 2.70\,$\mu$m with a 3.0\,$\mu$m operational band-edge goal. A reduction to 2.5\,$\mu$m remains the formal engineering off-ramp if thermal, detector, cooling, mass, power, or cost constraints require it. The \facility\ does not carry a mid-infrared channel. Coordinated ground-based mid-infrared telescopes provide the thermal fluxes required to infer diameter and albedo, while the space mission supplies contemporaneous reflected-light measurements and observing geometry. The program combines recovery, physical characterization, orbit refinement, and covariance-based hazard assessment. Its CODES dynamics system and OGFinder-to-OpenOrb processing path connect measured astrometry to reproducible orbit solutions and close-approach predictions.
}
\newcommand{\wpkeywords}{\textbf{Keywords:} near-Earth objects, planetary defense, small bodies, astrometry, orbit determination, optical and near-infrared spectroscopy}
\newcommand{\wpchapteroffset}{3}
\newcommand{\MakeVolumeBody}{%
  \definecolor{Blue1}{HTML}{C89562}%
  \definecolor{Blue2}{HTML}{A4612E}%
  \definecolor{Blue3}{HTML}{713F24}%
}

\newcommand{\MakeFrontCover}{%
  \begin{titlepage}
  \thispagestyle{empty}\sffamily\centering
  \noindent\colorbox{Blue3}{\parbox[t]{\dimexpr\textwidth-2\fboxsep\relax}{%
    \vspace{4mm}\centering
    {\color{white}\bfseries\fontsize{24}{29}\selectfont 3.5-meter Segmented-Mirror\\[1.5mm]
      Robotic Space Telescope}\\[3.5mm]
    {\color{white}\Large Mission White Paper}\\[1.5mm]
    {\color{white!88}\large \wpvolumelabel. \wpvolumetitle}%
    \vspace{4mm}}}
  \vfill
  \includegraphics[width=\textwidth]{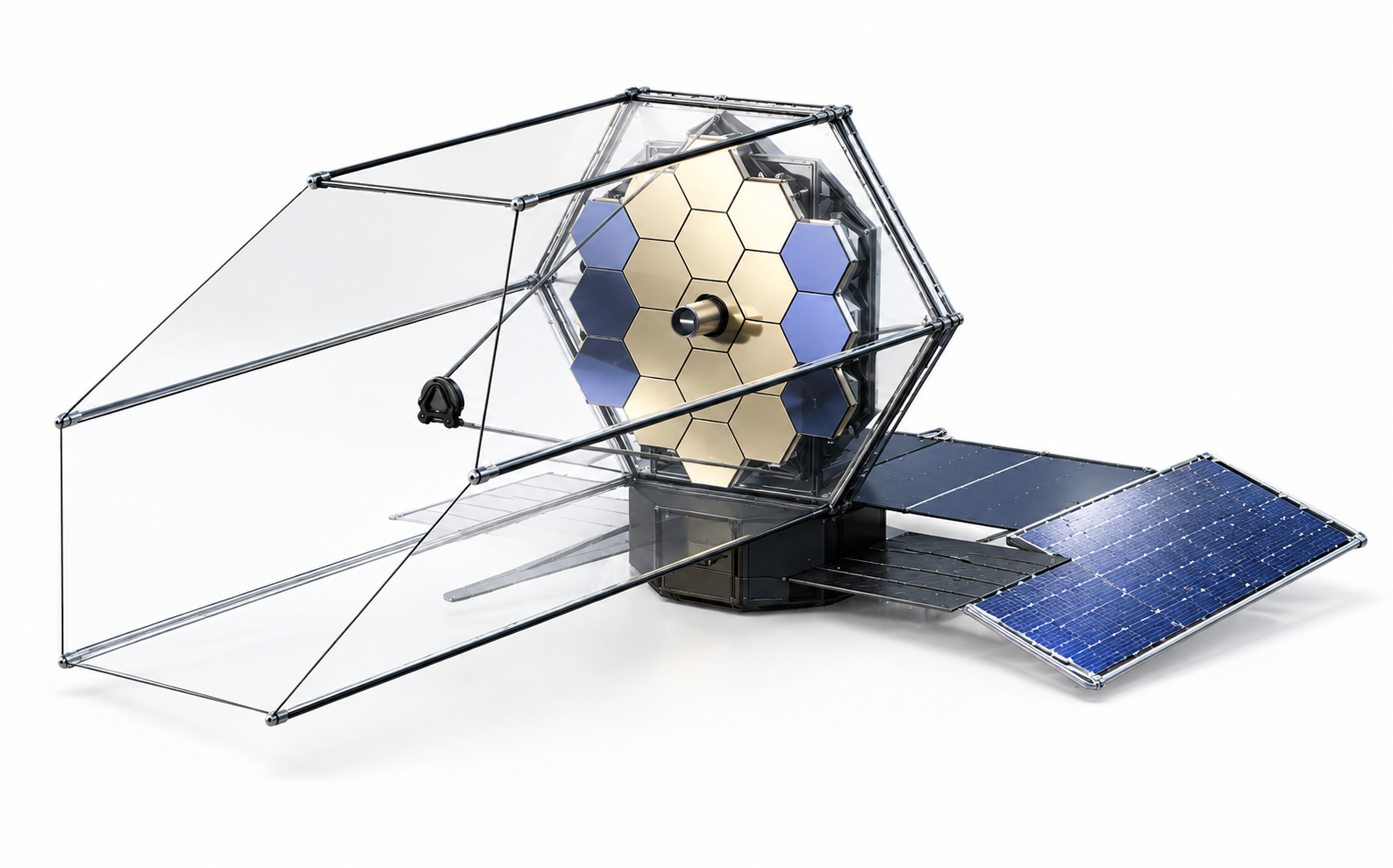}
  \vfill
  {\color{ink}\bfseries\normalsize \wpauthors\par}
  \vspace{3mm}
  {\color{muted}\small
    $^{1}$ Korea Institute for Advanced Study \textperiodcentered\
    $^{2}$ Korea Astronomy and Space Science Institute \\
    $^{3}$ University of Ulsan \textperiodcentered\
    $^{4}$ University of Science and Technology\\
    $^{5}$ Kyungpook National University \textperiodcentered\
    $^{6}$ Space Telescope Science Institute \textperiodcentered\
    $^{7}$ Pusan National University\\
    $^{8}$ Korea National University of Education \textperiodcentered\
    $^{9}$ Seoul National University \textperiodcentered\
    $^{10}$ Yonsei University\par}
  \vspace{5mm}
  {\color{Blue3}\bfseries\Large 2026}
  \vspace{4mm}
  \end{titlepage}}

\newcommand{\MakeColophon}{%
  \clearpage
  \thispagestyle{empty}\sffamily
  \null\vfill
  \noindent{\color{muted}\footnotesize
    {\color{ink}\bfseries 3.5-meter Segmented-Mirror Robotic Space Telescope}\\
    Mission White Paper: \wpvolumelabel. \wpvolumetitle\\[5mm]
    {\color{rose}\bfseries Release date September 17, 2026\\[5mm]}
    \textcopyright\ 2026 Korea Astronomy and Space Science Institute and the
    authors. All rights reserved.\\[2mm]
    Prepared by the mission study team at the Korea Astronomy and Space Science
    Institute, University of Ulsan, University of Science and Technology, the
    Space Telescope Science Institute, the Korea Institute for Advanced Study,
    Pusan National University, Korea National University of Education, Seoul
    National University, Kyungpook National University, and Yonsei University.\\[5mm]
    {\color{ink}\bfseries Suggested citation:} Kim, J., Kang, Y.-W., Lee, S.-H.,
    et al.\ (2026), \textit{3.5-meter Segmented-Mirror Robotic Space Telescope:
    Mission White Paper, \wpvolumelabel. \wpvolumetitle}.\\[2mm]
    {\color{ink}\bfseries Corresponding author:} Sang Hyun Lee
    \textperiodcentered\ \texttt{shlee@kasi.re.kr}.\par}
  \vspace{8mm}
  \clearpage}

\newcommand{\MakeBackCover}{%
  \clearpage
  \ifodd\value{page} \hbox{}\thispagestyle{empty}\newpage \fi
  \thispagestyle{empty}\null
  \clearpage
  \thispagestyle{empty}\sffamily\centering
  \null\vspace*{\stretch{1}}
  \includegraphics[width=\textwidth]{vol4_3.5ST_GPT.png}
  \vspace*{\stretch{2.2}}
  \noindent\colorbox{Blue3}{\parbox[t]{\dimexpr\textwidth-2\fboxsep\relax}{%
    \vspace{3mm}\centering
    {\color{white}\bfseries\Large 3.5-meter Segmented-Mirror Robotic Space Telescope}\\[1.5mm]
    {\color{white!88}\normalsize Mission White Paper: \wpvolumelabel. \wpvolumetitle}%
    \vspace{3mm}}}
  \vspace{3mm}
  \clearpage}

\hypersetup{
  colorlinks=true,
  linkcolor=blue,
  citecolor=teal,
  urlcolor=gold,
  pdftitle={3.5-meter Segmented-Mirror Robotic Space Telescope -- \wpvolumelabel. \wpvolumetitle},
  pdfauthor={Juhan Kim et al.}}

\begin{document}
\frontmatter
\MakeFrontCover
\MakeColophon
\thispagestyle{fancy}
\vspace*{0.5em}
\noindent{\Large\bfseries Abstract}\par\smallskip
\noindent\wpabstract\par\medskip
\noindent\wpkeywords
\clearpage
\tableofcontents

\mainmatter
\renewcommand{\chaptername}{Volume}
\renewcommand{\thechapter}{\Roman{chapter}}
\setcounter{chapter}{\wpchapteroffset}
\MakeVolumeBody
\def\APPENDIXCBOOK{1}
\ifdefined\APPENDIXCBOOK
\let\appendixCfinish\relax
\definecolor{orange}{HTML}{B55220}
\colorlet{cDeep}{white}
\colorlet{cPanel}{paper}
\colorlet{cPanelTwo}{panel}
\colorlet{cText}{ink}
\colorlet{cMuted}{muted}
\colorlet{cTeal}{teal}
\colorlet{cGold}{gold}
\colorlet{cOrange}{orange}
\colorlet{cRed}{rose}
\colorlet{cBlue}{blue}
\arrayrulecolor{line}
\newcolumntype{Z}{>{\centering\arraybackslash}X}
\newcolumntype{P}[1]{>{\raggedright\arraybackslash}p{#1}}
\newcolumntype{M}[1]{>{\raggedright\arraybackslash}m{#1}}
\newcolumntype{C}[1]{>{\centering\arraybackslash}m{#1}}
\newcommand{\topcell}[1]{%
  \begin{minipage}[t]{\linewidth}\vspace{0pt}#1\end{minipage}}
\newtcolorbox{resultbox}[1][]{
  enhanced,
  colback=paper,
  colframe=teal,
  coltext=ink,
  boxrule=0.9pt,
  arc=2mm,
  left=2.2mm,right=2.2mm,top=1.8mm,bottom=1.8mm,
  #1
}
\newtcolorbox{warningbox}[1][]{
  enhanced,
  colback=rose!4,
  colframe=rose,
  coltext=ink,
  boxrule=0.9pt,
  arc=2mm,
  left=2.2mm,right=2.2mm,top=1.8mm,bottom=1.8mm,
  #1
}
\newcommand{\code}[1]{\texttt{\color{teal}#1}}
\chapter{Solar-System Small Bodies and Planetary Defense}
\label{app:solar-system}
\else
\documentclass[10pt,onecolumn]{article}

\usepackage[a4paper,margin=18mm,top=17mm,bottom=18mm]{geometry}
\raggedbottom
\renewcommand{\topfraction}{0.9}
\renewcommand{\floatpagefraction}{0.85}
\usepackage{microtype}
\usepackage{amsmath,amssymb}
\usepackage{siunitx}
\usepackage{booktabs}
\usepackage{tabularx}
\usepackage{array}
\usepackage{longtable}
\usepackage{graphicx}
\usepackage[dvipsnames,svgnames,table]{xcolor}
\usepackage[most]{tcolorbox}
\usepackage{titlesec}
\usepackage{enumitem}
\usepackage{fancyhdr}
\usepackage{hyperref}
\usepackage{caption}
\usepackage{float}

\definecolor{ink}{HTML}{151B23}
\definecolor{titlebg}{HTML}{713F24}
\definecolor{softink}{HTML}{263441}
\definecolor{muted}{HTML}{5C6875}
\definecolor{paper}{HTML}{F6F7F2}
\definecolor{panel}{HTML}{FFFFFF}
\definecolor{line}{HTML}{D8DEE5}
\definecolor{teal}{HTML}{007C77}
\definecolor{blue}{HTML}{A4612E}
\definecolor{gold}{HTML}{B86B00}
\definecolor{rose}{HTML}{A7354D}
\definecolor{green}{HTML}{23724A}
\definecolor{orange}{HTML}{B55220}

\colorlet{cDeep}{white}
\colorlet{cPanel}{paper}
\colorlet{cPanelTwo}{panel}
\colorlet{cText}{ink}
\colorlet{cMuted}{muted}
\colorlet{cTeal}{teal}
\colorlet{cGold}{gold}
\colorlet{cOrange}{orange}
\colorlet{cRed}{rose}
\colorlet{cBlue}{blue}

\arrayrulecolor{line}
\hypersetup{
  colorlinks=true,
  linkcolor=blue,
  citecolor=teal,
  urlcolor=gold,
  pdftitle={Volume IV: Solar-System Small Bodies and Planetary Defense},
  pdfauthor={Juhan Kim}
}

\pagestyle{fancy}
\fancyhf{}
\fancyhead[L]{\sffamily\footnotesize 3.5-meter Segmented-Mirror Robotic Space Telescope}
\fancyhead[R]{\sffamily\footnotesize Volume IV: Solar-System Science}
\fancyfoot[C]{\sffamily\footnotesize IV-\thepage}
\renewcommand{\headrulewidth}{0.35pt}

\titleformat{\section}{\sffamily\Large\bfseries\color{ink}}{\thesection}{0.6em}{}
\titleformat{\subsection}{\sffamily\large\bfseries\color{blue}}{\thesubsection}{0.6em}{}
\titleformat{\subsubsection}{\sffamily\normalsize\bfseries\color{softink}}{\thesubsubsection}{0.6em}{}
\titlespacing*{\section}{0pt}{1.1em}{0.45em}
\titlespacing*{\subsection}{0pt}{0.85em}{0.3em}
\captionsetup{font={small,sf},labelfont={bf,color=blue}}
\setlist[itemize]{leftmargin=*,itemsep=2pt,topsep=3pt}
\setlist[enumerate]{leftmargin=*,itemsep=2pt,topsep=3pt}
\renewcommand{\arraystretch}{1.18}
\renewcommand{\thesection}{IV.\arabic{section}}
\renewcommand{\thesubsection}{IV.\arabic{section}.\arabic{subsection}}
\renewcommand{\thesubsubsection}{IV.\arabic{section}.\arabic{subsection}.\arabic{subsubsection}}
\renewcommand{\thefigure}{IV.\arabic{figure}}
\renewcommand{\thetable}{IV.\arabic{table}}
\renewcommand{\theequation}{IV.\arabic{equation}}
\newcolumntype{Y}{>{\raggedright\arraybackslash}X}
\newcolumntype{Z}{>{\centering\arraybackslash}X}
\newcolumntype{P}[1]{>{\raggedright\arraybackslash}p{#1}}
\newcolumntype{M}[1]{>{\raggedright\arraybackslash}m{#1}}
\newcolumntype{C}[1]{>{\centering\arraybackslash}m{#1}}
\newcommand{\topcell}[1]{%
  \begin{minipage}[t]{\linewidth}\vspace{0pt}#1\end{minipage}}
\emergencystretch=3em

\newtcolorbox{resultbox}[1][]{
  enhanced,
  colback=paper,
  colframe=teal,
  coltext=ink,
  boxrule=0.9pt,
  arc=2mm,
  left=2.2mm,right=2.2mm,top=1.8mm,bottom=1.8mm,
  #1
}
\newtcolorbox{warningbox}[1][]{
  enhanced,
  colback=rose!4,
  colframe=rose,
  coltext=ink,
  boxrule=0.9pt,
  arc=2mm,
  left=2.2mm,right=2.2mm,top=1.8mm,bottom=1.8mm,
  #1
}
\newcommand{\facility}{3.5-meter Segmented-Mirror Robotic Space Telescope}
\newcommand{\code}[1]{\texttt{\color{teal}#1}}
\newcommand{\appendixCfinish}{\end{document}}

\begin{document}
\thispagestyle{empty}
\begin{tcolorbox}[enhanced,colback=titlebg,colframe=titlebg,arc=2.5mm,
  left=5mm,right=5mm,top=7mm,bottom=7mm]
{\sffamily\color{white}\fontsize{23}{28}\selectfont\bfseries
Volume IV: Solar-System Small Bodies and Planetary Defense}\\[2.2mm]
{\sffamily\color{white}\fontsize{14}{18}\selectfont on the \facility}\\[3mm]
{\sffamily\color{white!78}\normalsize NEO recovery, optical and
near-infrared characterization, orbit propagation, and close-approach
validation}
\end{tcolorbox}
\phantomsection
\label{app:solar-system}
\fi

\vspace{2mm}
\noindent
\begin{tabularx}{\textwidth}{@{}P{31mm}Y@{}}
\topcell{\bfseries\color{cGold}Proposal role} &
\topcell{Planetary defense and small-body population science with a clear
separation between discovery and post-discovery characterization.}\\
\topcell{\bfseries\color{cGold}Baseline claim} &
\topcell{High-value follow-up, physical characterization, and orbit refinement
are feasible within the present optical and near-infrared concept.}\\
\topcell{\bfseries\color{cGold}Coordination claim} &
\topcell{Targeted ground-based MIR observations provide thermal fluxes.
The \facility\ provides optical and near-infrared photometry, spectra,
astrometry, and rotational phase information.}\\
\topcell{\bfseries\color{cGold}Feasibility \& Scope} &
\topcell{Performance is stated as a requirement or a demonstrated comparison.
Unverified limiting magnitudes and completeness claims are not treated as
mission performance.}
\end{tabularx}

\section{Science Scope}

Recent space-observatory studies have established the value of linking
discovery, rapid follow-up, and physical interpretation across observing
facilities \cite{roy2026IV,wevers2026IV}. This volume applies that network model
to small bodies. The 3.5ST supplies repeated optical and near-infrared
astrometry, spectra, and light curves. Ground facilities supply targeted
mid-infrared thermal measurements because the 3.5ST baseline does not include
an MIR channel. The resulting data products are orbit covariance, size and
albedo constraints, and physically interpretable warning intervals rather
than a generic transient-alert capability.

The original proposal combines NEO discovery, physical characterization,
orbit refinement, impact-risk analysis, and early warning in one program.
The scientific sequence is coherent although each stage imposes different
hardware and operational requirements. The mission architecture therefore
separates the program into four data products.

\begin{enumerate}[leftmargin=7mm,itemsep=2mm]
\item \textbf{\color{cTeal}Recovery and astrometry.} Measure accurate
positions of recently discovered objects in the International Celestial
Reference Frame (ICRF). The ICRF is the quasar-defined inertial reference
frame used for high-accuracy celestial astrometry. The new positions extend
the observational arc before the ephemeris uncertainty grows.
\item \textbf{\color{cGold}Physical characterization.} Measure colors,
spectra, light curves, and phase behavior. Combine the measurements with
thermal flux from coordinated ground-based MIR observations.
\item \textbf{\color{cOrange}Dynamical refinement.} Fit gravitational and
non-gravitational accelerations, including Yarkovsky drift when the data arc
and physical priors support the fit.
\item \textbf{\color{cRed}Hazard assessment.} Propagate the full orbit
covariance through close encounters. A nominal trajectory alone is not an
impact probability.
\end{enumerate}

The \facility\ provides its greatest gain through its aperture, angular
resolution, robotic scheduling, and stable space environment. A dedicated
discovery survey requires the additional capabilities of large étendue,
repeated cadence, moving-object processing, and access toward the Sun.

\section{Ground-Based and Space-Based NEO Survey Regimes}
\label{sec:ground-space-surveys}

Ground and space surveys serve complementary roles. Wide-field ground
observatories deliver large optical étendue at comparatively low cost.
Operators can replace or upgrade their instruments and distribute longitude
coverage across a network. Rubin Observatory demonstrates the resulting
survey power through repeated optical imaging over a very large field
\cite{Ivezic2019}. A space survey should therefore not be justified as a
smaller copy of a ground survey. Its scientific case rests on access to a
different observational phase space. An atmosphere-free platform provides
stable optical and near-infrared calibration, a larger usable time window,
and a field of regard that can be engineered closer to the Sun. The
coordinated ground segment provides the MIR measurement because the 3.5ST
payload does not include an MIR channel.

\subsection{Atmosphere, airmass, and the usable sky window}

For a ground observation at zenith angle $z$ the airmass is approximately
$X\simeq\sec z$ away from the horizon. Atmospheric extinction contributes
$A_\lambda=k_\lambda X$. Increasing airmass attenuates the source and
increases wavelength-dependent extinction, differential refraction, seeing,
and calibration systematics. A moving-object survey consequently
imposes a maximum acceptable airmass rather than using the full geometric
hemisphere. Daylight and twilight, the horizon, weather, lunar background,
and seasonal visibility further reduce the effective sky and cadence window.
The combined restrictions narrow the ground-based sky window. The excluded
region is set by data-quality and scheduling requirements, not only by
whether the target is geometrically above the horizon.

Mid-infrared astronomy from the ground uses restricted atmospheric windows
at exceptional dry sites. ESO/VISIR provides one established example through
the M, N, and Q windows near 5, 10, and 19\,$\mu$m. The warm atmosphere and
telescope produce a high photon background. Ground observations subtract
the background through chopping, nodding, and short targeted integrations
\cite{ESOVISIR2019}. The resulting mode does not support a continuous
all-sky MIR census. Targeted observing instead provides thermal
characterization of selected targets with sufficiently accurate ephemerides.
The adopted division of
responsibility assigns this targeted role to ground-based facilities.

\begin{table}[H]
\centering
\caption{Division of NEO responsibilities between ground facilities and
the 3.5ST space observatory. MIR denotes targeted ground-based follow-up and
does not denote a 3.5ST observing mode.}
\small
\begin{tabularx}{\textwidth}{@{}P{30mm}YY@{}}
\toprule
\textbf{\color{cGold}Survey property} &
\textbf{\color{cGold}Ground surveys and MIR facilities} &
\textbf{\color{cGold}3.5ST optical and near-infrared program}\\
\midrule
Primary strength &
Large optical étendue, long temporal baselines, upgradeable instruments,
and targeted MIR thermal measurements &
Atmosphere-free stability, rapid scheduling, precise astrometry, and
engineered solar-elongation access\\
\addlinespace
Usable wavelength range &
Visible and near-infrared survey bands plus restricted MIR atmospheric
windows. Transmission and sky background vary with water vapor and airmass &
The baseline 0.2--1.5\,$\mu$m range, with calibrated throughput to
2.70\,$\mu$m and a 3.0\,$\mu$m band-edge goal under study. No MIR detector or
MIR survey mode is included\\
\addlinespace
Sky and time window &
Night, horizon, airmass, weather, Moon, twilight, and season constrain each
field &
No weather, horizon, or day--night cycle. Sun/Earth/Moon avoidance and
spacecraft thermal limits define the field of regard\\
\addlinespace
Image and calibration &
Seeing, extinction, telluric absorption, and differential refraction vary
with time and direction &
Stable PSF and throughput are achievable. Detector persistence, cosmic
rays, pointing, and calibration must still be controlled\\
\addlinespace
NEO selection &
Optical discovery remains biased against low-albedo objects. Targeted MIR
flux constrains diameter after an ephemeris becomes available &
Reflected-light detection has the same albedo dependence. Space access and
stable calibration improve recovery but do not remove the diameter and
albedo degeneracy\\
\addlinespace
Cadence and recovery &
Weather and daylight can interrupt tracklets. A longitude-distributed
network can mitigate gaps &
Repeated cadence can continue across terrestrial day and weather, subject
to field-of-regard, downlink, and mission-scheduling constraints\\
\addlinespace
Lifecycle and risk &
Accessible maintenance, detector replacement, and incremental expansion &
Higher cost and launch risk, finite consumables and telemetry, and usually
limited servicing after launch\\
\bottomrule
\end{tabularx}
\label{tab:ground-space-neo}
\end{table}

\clearpage
\subsection{Space-observatory advantages relevant to the \facility}

\begin{enumerate}[leftmargin=7mm,itemsep=2mm]
\item \textbf{\color{cTeal}Stable optical and near-infrared characterization.}
The atmosphere-free measurements provide repeatable colors, spectra, phase
curves, and astrometry. The data constrain taxonomy, rotation, and the
reflected-light state at the epoch of a ground-based MIR observation.
\item \textbf{\color{cGold}Larger effective observing window.}
The absence of weather, daylight, horizon, and airmass restrictions supports
uniform revisit timing and more reliable tracklet linking. The gain must be
quantified with the actual Sun/Earth/Moon avoidance geometry rather than
described as unrestricted all-sky access.
\item \textbf{\color{cOrange}Access toward the Sun.}
A thermally qualified baffle and sunshade can observe smaller solar
elongations than night-time ground surveys, improving sensitivity to objects
approaching from directions that are poorly sampled in optical surveys
\cite{Mainzer2023}.
\item \textbf{\color{cBlue}Stable astrometry and photometry.}
Removing seeing, atmospheric dispersion, temporal extinction variations, and telluric
structure improves repeatability. The remaining floor is set by PSF
calibration, pointing reconstruction, detector effects, and the inertial
reference catalog.
\item \textbf{\color{cRed}Complementarity with ground assets.}
The program assigns complementary measurements to ground and space
facilities. Ground surveys provide optical étendue, long temporal baselines,
radar constraints, and MIR thermal flux. The space mission provides
weather-independent optical and near-infrared follow-up, precise astrometry,
and rapid scheduling.
\end{enumerate}

\begin{resultbox}
\textbf{\color{cGold}Mission consequence.}
The \facility\ does not claim MIR discovery or MIR sensitivity. The mission
provides rapid recovery, stable reflected-light characterization, and orbit
refinement. A ground-based MIR partner supplies the independent
thermal measurement needed to infer diameter and albedo.
\end{resultbox}

\section{Size, Albedo, and Coordinated Thermal Follow-up}

For an asteroid with absolute magnitude $H$ and visible geometric albedo
$p_V$, the standard optical diameter relation \cite{RyanWoodward2010} is
\begin{equation}
  D\,[\mathrm{m}]
  =1.329\times10^6\,p_V^{-1/2}10^{-0.2H}.
  \label{eq:h-diameter}
\end{equation}
The absolute magnitude $H$ specifies the brightness at the standard geometry
$r=\Delta=1$\,au and zero phase angle. For a real observation the apparent
visual magnitude and the corresponding diameter threshold follow
\cite{Bowell1989}
\begin{equation*}
\begin{aligned}
 m_V &=
 H+5\log_{10}(r\Delta)-2.5\log_{10}\Phi(\alpha),\\
 D_{\min}\,[\mathrm{m}] &=
 1.329\times10^6\,p_V^{-1/2}10^{-0.2m_{\rm lim}}
 \frac{r\Delta}{\sqrt{\Phi(\alpha)}} ,
\end{aligned}
\end{equation*}
where $r$ is the heliocentric distance, $\Delta$ is the distance between the
telescope and the NEO in au, and $\Phi(\alpha)$ is the normalized phase
function.
At fixed albedo and limiting magnitude the detectable minimum
diameter grows linearly with both $r$ and $\Delta$ and increases at
unfavorable phase angle. For example at $m_{\rm lim}=25$, $p_V=0.14$,
$r=1$\,au, and $\Phi=1$, the corresponding thresholds are 0.36, 3.6, and
36\,m for $\Delta=0.01$, 0.1, and 1\,au, respectively.
The equation shows why reflected-light discovery does not determine impact
energy or the resulting hazard level. A large dark body may have the same
$H$ as a smaller reflective body. Ground-based thermal-infrared flux breaks
much of the degeneracy and supports a diameter estimate when the observing
geometry and thermal model are known \cite{Mainzer2011,Mainzer2023}.

\begin{figure}[H]
\centering
\includegraphics[width=0.96\textwidth]{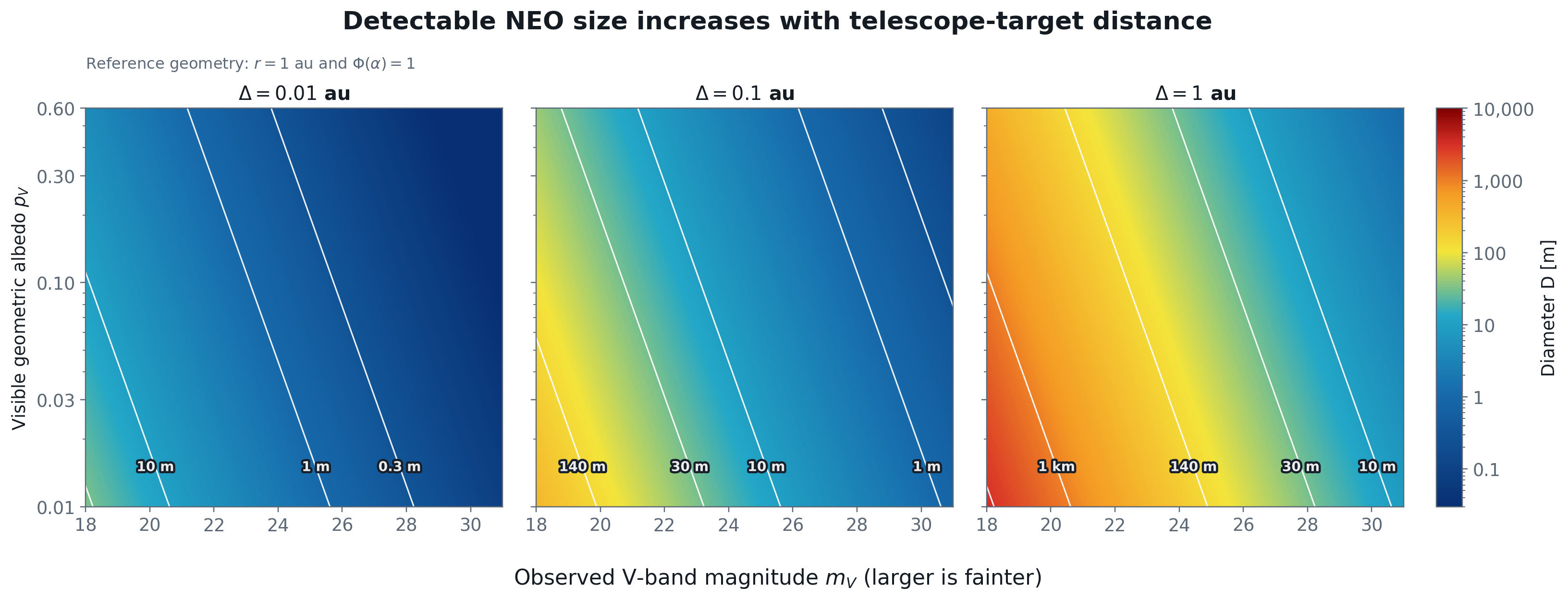}
\caption{Distance-aware optical detection threshold. Each panel gives the
diameter that produces observed magnitude $m_V$ as a function of visible
geometric albedo $p_V$ at distance $\Delta$ from the telescope. The shared color
scale and continuous white contours give diameter $D$. The reference
calculation sets $r=1$\,au and $\Phi(\alpha)=1$. The color at a survey's
$m_V=m_{\rm lim}$ gives an optimistic minimum detectable diameter. General
geometry scales the result by $r/\sqrt{\Phi(\alpha)}$. The figure presents a
flux threshold relation rather than a survey-completeness prediction.}
\label{fig:neo-h-albedo-diameter}
\end{figure}

\subsection{Why ground-based mid-infrared follow-up is required}

The 3.5ST reflected-light data alone do not provide an independent size
measurement. The reflected flux constrains a combination proportional to
$p_VD^2$. Equation~\ref{eq:h-diameter} leaves the diameter uncertain by
$p_V^{-1/2}$. A ground-based MIR flux measurement adds the thermal constraint.
The Near-Earth Asteroid Thermal Model gives the schematic dependence
\begin{equation}
 F_{\nu}^{\rm th}
 \mathrel{\propto}
 \frac{\epsilon D^2}{\Delta^2}
 B_\nu\!\left(T\right),
 \qquad
 T\mathrel{\propto}
 \left(\frac{1-A}{\eta\epsilon r^2}\right)^{1/4},
 \label{eq:thermal-diameter}
\end{equation}
where $\Delta$ and $r$ are the observer and heliocentric distances. The
dimensionless quantity $\epsilon$ is the bolometric thermal emissivity. The
emissivity multiplies the greybody radiance and enters the radiative-equilibrium
temperature as $T\propto\epsilon^{-1/4}$. The dimensionless NEATM beaming
parameter $\eta$ empirically rescales the surface-temperature distribution
and represents the combined effects of thermal inertia, rotation, surface
roughness, and observing geometry. Increasing $\eta$ lowers the modeled
temperature as $T\propto\eta^{-1/4}$ \cite{Harris1998}. A joint fit must
either constrain $\epsilon$ and $\eta$ from the thermal data or propagate
their prior uncertainty into the diameter posterior.
The diameter enters the emitted flux through the projected area as $D^2$.
The Bond albedo $A$ enters only through a fourth-root temperature correction
\cite{Harris1998,Mainzer2011}. A measured thermal spectral energy distribution
therefore provides a nearly albedo-independent diameter constraint. The
thermal constraint does not require an assumed visible geometric albedo $p_V$
although shape, thermal inertia, surface roughness, rotation, and the beaming
parameter $\eta$ remain model uncertainties. Combining a ground-based MIR
diameter with the 3.5ST value of $H$ determines $p_V$ rather than assuming
$p_V$. The distinction is essential for hazard assessment because impact
energy scales approximately as $D^3$.

Wien's displacement law gives
\begin{equation}
  \lambda_{\rm peak}
  =\frac{b}{T},
  \qquad
  b\simeq2898\,\mu\mathrm{m\,K}.
  \label{eq:wien}
\end{equation}
The left-hand side $\lambda_{\rm peak}$ is the wavelength at which the
blackbody spectral radiance per unit wavelength $B_\lambda(T)$ reaches its
maximum. The symbol $T$ is the effective blackbody temperature in kelvin and
$b$ is Wien's displacement constant. Equation~\ref{eq:wien} returns
$\lambda_{\rm peak}$ in micrometers when $b$ is expressed in
$\mu\mathrm{m\,K}$. The numerical constant applies to the peak of
$B_\lambda$. A spectrum expressed per unit frequency as $B_\nu$ has a
different numerical peak.
Objects at 250--400\,K peak near 11.6--7.2\,$\mu$m. The 3.5ST
0.2--1.5\,$\mu$m range measures reflected sunlight for most NEOs and does not
sample the thermal maximum. Ground-based M-band and N-band observations
sample the warm continuum through the available atmospheric windows. NEO
Surveyor provides a space-based reference that demonstrates the required
thermal wavelength coverage \cite{Mainzer2023}. Its detector bands do not
define 3.5ST requirements.

\begin{figure}[H]
\centering
\includegraphics[width=0.96\textwidth]{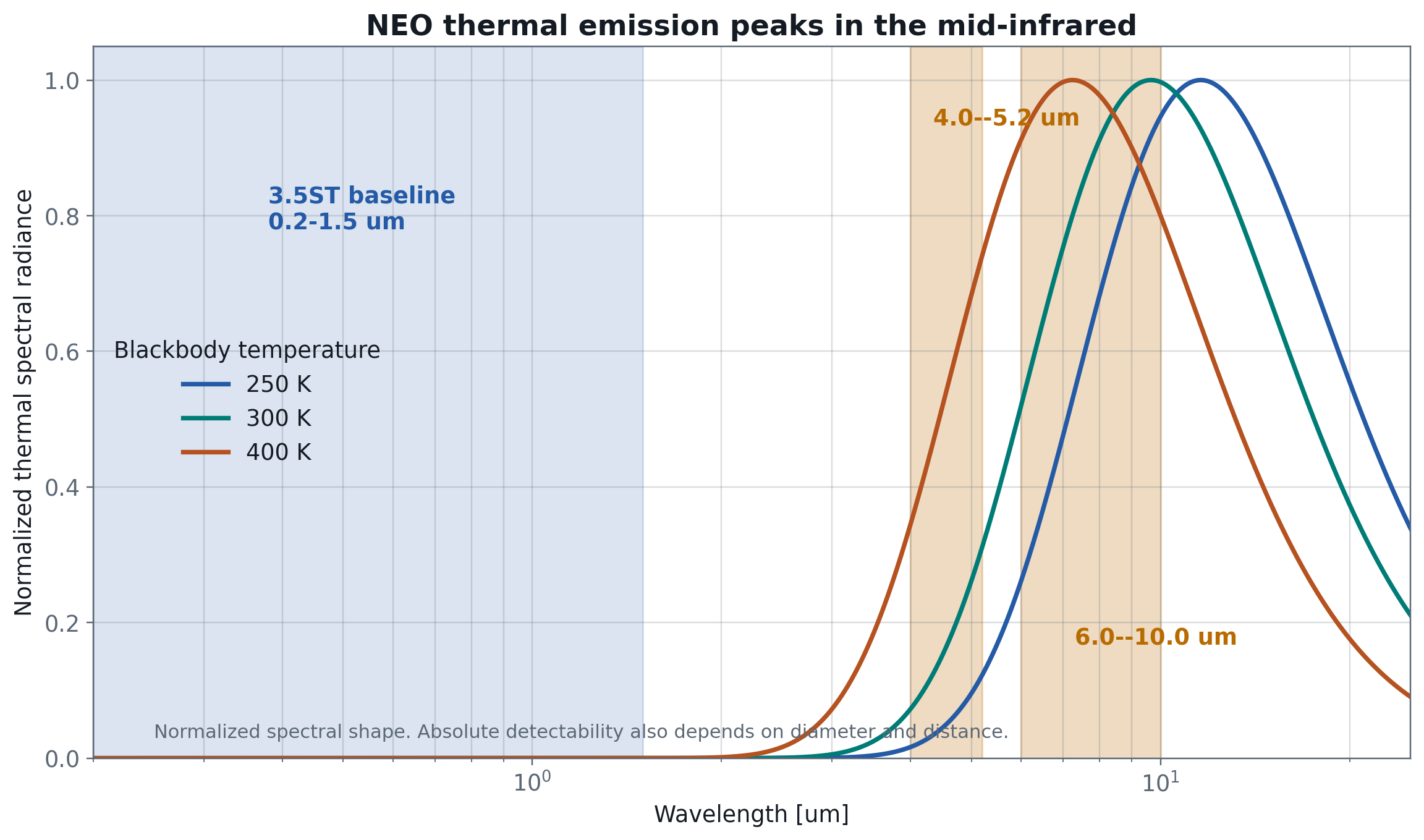}
\caption{Normalized blackbody shapes for representative NEO temperatures.
The gold intervals mark the NEO Surveyor channel ranges as a wavelength
reference. The \facility\ does not use the intervals. Ground-based MIR
facilities sample portions of the same thermal continuum through their
available atmospheric windows.}
\end{figure}

\begin{warningbox}
\textbf{\color{cOrange}Instrument boundary.}
MIR observations are assigned to ground-based telescopes. No MIR detector,
cold MIR fore-optics, MIR exposure-time calculation, or MIR discovery yield
belongs to the 3.5ST baseline. This volume uses thermal physics only to
define the external data needed for diameter and albedo inference.
\end{warningbox}

\subsection{Ground-based MIR follow-up interface}

The program obtains the thermal measurement as an external data product.
The 3.5ST observation determines the ICRF position, the optical and
near-infrared spectral energy distribution, the rotational phase, and the
phase-angle correction. A ground-based MIR telescope measures calibrated
thermal flux in one or more available atmospheric windows. CODES evaluates
the heliocentric and observer distances at both epochs. A joint NEATM or
thermophysical fit then determines diameter, visible albedo, and their
covariance.

\begin{table}[H]
\centering
\caption{Data interface for coordinated physical characterization. The
MIR entry denotes a ground-based measurement and does not consume 3.5ST
observing time.}
\begin{tabularx}{\textwidth}{@{}P{34mm}P{47mm}Y@{}}
\toprule
\textbf{\color{cGold}Contributor} &
\textbf{\color{cGold}Required product} &
\textbf{\color{cGold}Role in the joint inference}\\
\midrule
3.5ST &
ICRF astrometry, 0.2--1.5\,$\mu$m fluxes, spectrum, light curve, and UTC
mid-exposure times &
Constrains $H$, taxonomy, rotation, phase behavior, and reflected flux\\
Ground-based MIR telescope &
Calibrated thermal flux, filter response, atmospheric transmission,
photometric uncertainty, and UTC mid-exposure time &
Constrains thermal emission and the diameter scale\\
CODES &
Observer geometry, heliocentric distance, phase angle, and orbit covariance
at every exposure &
Computes the geometric corrections for the optical and thermal measurements\\
Joint physical model &
NEATM or thermophysical likelihood with rotational and calibration
uncertainties &
Determines posterior distributions for $D$, $p_V$, $\eta$, and derived impact
energy\\
\bottomrule
\end{tabularx}
\label{tab:ground-mir-interface}
\end{table}

The ground observation should occur as close as practical to the 3.5ST
light-curve sequence. Non-simultaneous measurements require a rotational
phase model and an uncertainty term for the evolving geometry. The collaboration
must exchange calibrated fluxes and filter curves rather than catalog
diameters alone. The retained observables allow the physical model to be
recomputed after a revised orbit, thermal model, or calibration becomes
available.

\begin{resultbox}
\textbf{\color{cTeal}Scope of the mission ETC.}
The 3.5ST exposure-time calculator covers only the adopted optical and
near-infrared instruments. Each ground MIR facility evaluates its own
sensitivity with measured atmospheric transmission, thermal background,
detector properties, and observing mode. MIR sensitivity and MIR yield are
not quoted as 3.5ST performance.
\end{resultbox}

\section{Baseline Optical and Near-Infrared Program}

\subsection{Astrometry and recovery}

For a well-sampled point-spread function, a useful photon-limit estimate is
\begin{equation}
  \sigma_{\rm cent}
  \simeq \frac{\mathrm{FWHM}}{2.355\,\mathrm{S/N}}.
  \label{eq:centroid}
\end{equation}
Equation~\ref{eq:centroid} estimates only the photon-limited centroid error.
The absolute astrometric error also depends on geometric-distortion stability,
detector effects, timing, guide-star errors, chromaticity, source trailing,
and the reference catalog. The mission requirement should therefore specify
both centroid precision and an externally validated ICRF floor.

\begin{figure}[H]
\centering
\includegraphics[width=0.94\textwidth]{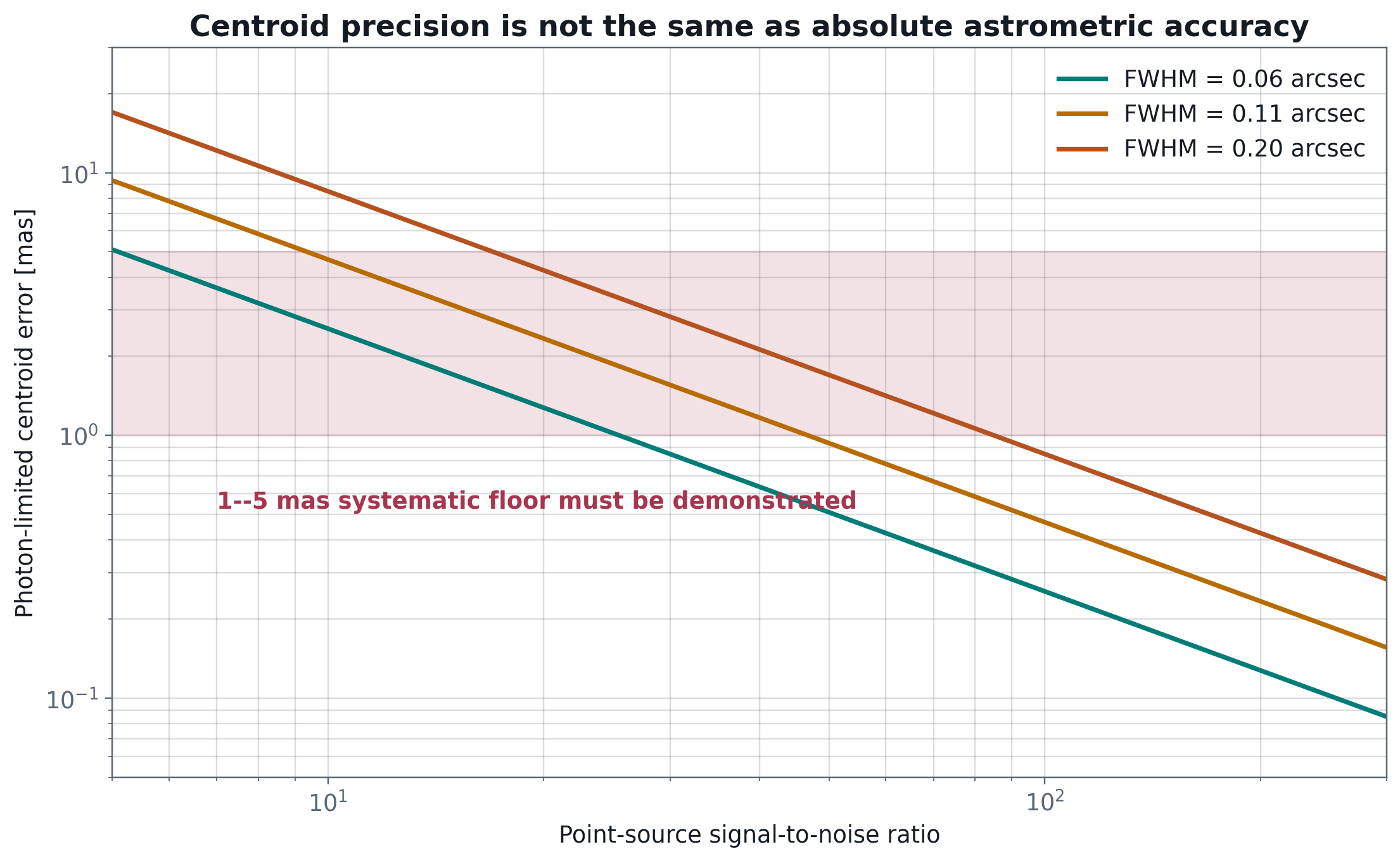}
\caption{Photon-limited centroid precision for three representative PSF
widths. The shaded floor is a calibration target, not a demonstrated
performance value.}
\end{figure}

The rapid Target-of-Opportunity mode acquires several short exposures that
limit trailing. A second sequence after a geometry-dependent interval extends
the orbital arc. The resulting astrometry includes the complete timing and
distortion solution. Synthetic moving-object injection must determine the
final cadence because a static-source survey does not measure the relevant
recovery efficiency.

\subsection{Composition, rotation, and phase behavior}

The Bus--DeMeo system uses spectra from 0.45 to 2.45\,$\mu$m and defines
24 taxonomic classes \cite{DeMeo2009}. The baseline 0.2--1.5\,$\mu$m
spectrograph therefore provides a useful but incomplete taxonomic interval.
Extension to 2.45\,$\mu$m would improve continuity with the established
classification system. Time-resolved photometry measures the rotation period,
light-curve amplitude, and evolving viewing geometry. The measurements
constrain the shape and spin priors used by thermal and Yarkovsky models.

\section{L2 Field of Regard}

At Sun--Earth L2 the Sun, Earth, and Moon occupy nearly the same sky
direction. The angular radius of the solar disk does not set the operational
exclusion angle. L2 remains the preferred orbit architecture. MEO is the
current fallback if launch-vehicle or programmatic constraints prevent L2;
LEO is not the planning baseline because of its less stable thermal and
observing environment for a long-duration survey.
Scattered light, sunshade geometry, thermal stability, Earth thermal
emission, lunar light, and allowed spacecraft attitudes set a much larger
keep-out region. The mission-wide schedule adopts a minimum solar elongation
of $90^\circ$. The $60^\circ$ value below is retained only as a relaxed
engineering study, and the $45^\circ$--$60^\circ$ region is not included in
the nominal time allocation.

For a fixed target at ecliptic latitude $\beta$ and solar exclusion angle
$\theta$ the purely geometric annual visibility fraction is
\begin{equation}
 f_{\rm vis}(\beta)=
 \begin{cases}
  1-\dfrac{1}{\pi}
  \cos^{-1}\!\left(\dfrac{\cos\theta}{\cos\beta}\right),
  & |\beta|<\theta,\\[2mm]
  1, & |\beta|\ge\theta.
 \end{cases}
 \label{eq:visibility}
\end{equation}
For the relaxed study value $\theta=60^\circ$, Equation~\ref{eq:visibility}
gives 243.5 observable days per year on the ecliptic and full-year access
above $|\beta|=60^\circ$. The visibility fractions describe geometry only;
the database-driven schedule uses $90^\circ$--$180^\circ$ for every nominal
program.

\begin{figure}[H]
\centering
\includegraphics[width=\textwidth]{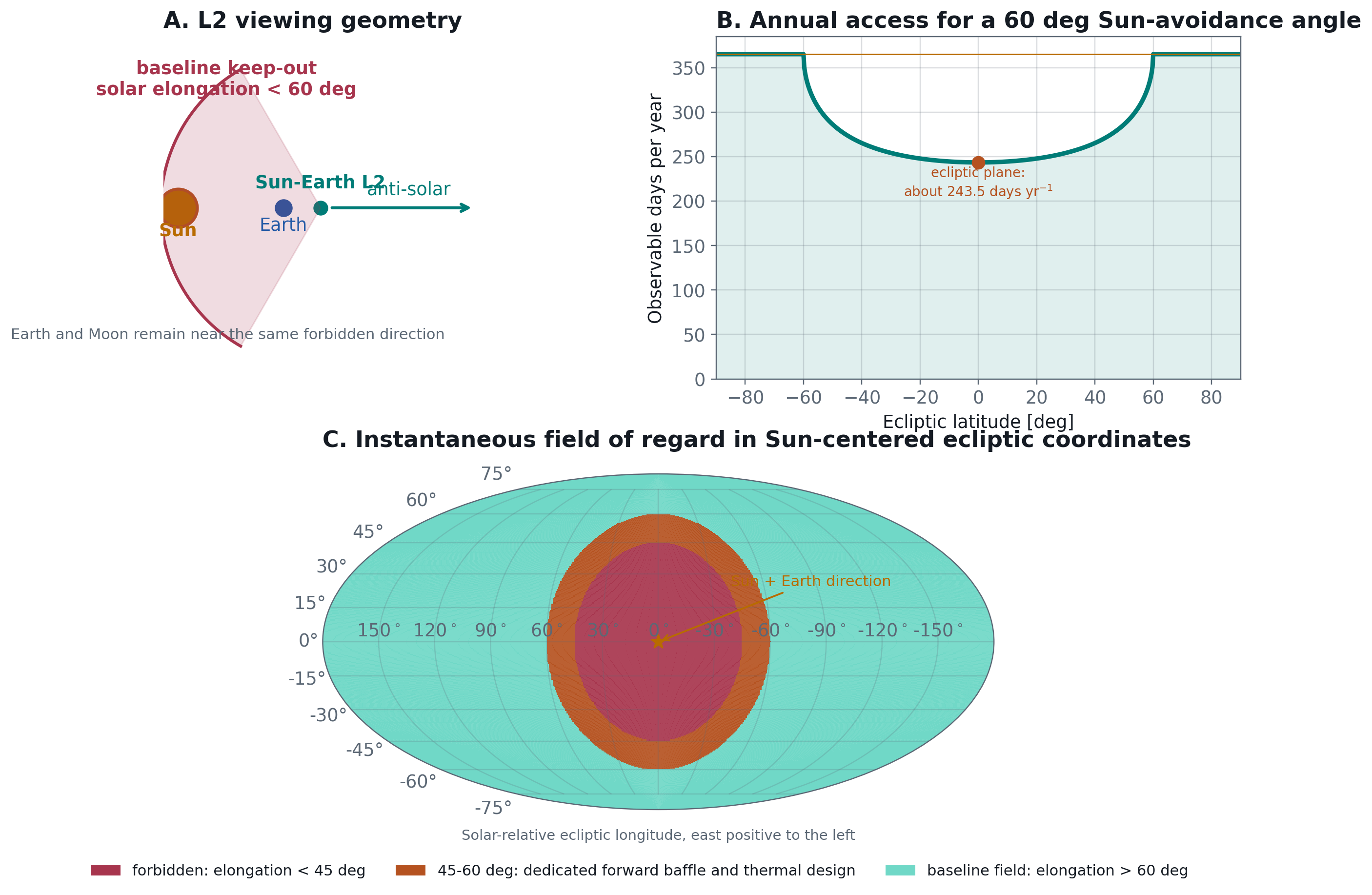}
\caption{L2 field-of-regard concept. Red marks a $45^\circ$ inner exclusion
zone. Orange marks the 45--$60^\circ$ region that requires a dedicated
forward baffle and thermal design. Teal marks the relaxed study region outside
$60^\circ$; the nominal scheduler begins at $90^\circ$. Earth albedo is suppressed by the near-new-Earth phase as seen
from L2 although Earth and Moon stray light, thermal emission, halo-orbit
offsets, and sunshade geometry still require a measured rejection model.}
\end{figure}

NEO discovery benefits from observing at small elongation because Atira and
other interior or Earth-approaching populations spend substantial time close
to the Sun on the sky. A reduction from $60^\circ$ to $45^\circ$ would increase
sensitivity to these populations. Reaching $45^\circ$ requires corresponding
designs for the baffle, sunshade, radiator view, detector background, and
allowed roll angles. Zodiacal emission also rises toward the ecliptic and
toward the Sun.
The exposure-time model must use a wavelength-dependent zodiacal map rather
than one scalar sky level.

\section{Observation Architecture and Feasibility}

The observing architecture assigns a measurable product and an acceptance
test to each science claim. Product-specific acceptance tests prevent a
photon-limited sensitivity estimate from being interpreted as a demonstrated
recovery rate or survey completeness.

\begin{longtable}{@{}P{30mm}P{36mm}P{46mm}P{46mm}@{}}
\caption{Product, observable, baseline status, and acceptance test for
each characterization step in the coordinated recovery program.}
\label{tab:c3-acceptance-tests}\\
\toprule
\color{cGold}\textbf{Product} &
\color{cGold}\textbf{Observable} &
\color{cGold}\textbf{Baseline status} &
\color{cGold}\textbf{Acceptance test}\\
\midrule
\endhead
Recovery &
ICRF position and angular rate &
Feasible with direct imaging and rapid scheduling &
Blind injection of trailed sources through detector, distortion, timing,
and catalog calibration\\
Rotation &
Time-series flux and color &
Feasible for targets with adequate cadence and S/N &
Recover period and amplitude from synthetic irregular windows\\
Taxonomy &
0.45--1.5\,$\mu$m spectrum at baseline &
Partial Bus--DeMeo interval &
Class confusion matrix versus magnitude, phase angle, and trailing\\
Diameter and albedo &
Reflected plus thermal flux &
Requires coordinated ground-based MIR data &
Joint fit to synthetic thermal models with shape and phase uncertainty\\
Yarkovsky &
Long-arc transverse acceleration $A_2$ &
Target dependent &
Recovery of injected $A_2$ after joint optical, radar, and thermal fitting\\
Discovery completeness &
Tracklets, linkage, selection function &
Not established for the baseline &
End-to-end survey simulation with field of regard, cadence, trailing,
confusion, and false-link control\\
\bottomrule
\end{longtable}

\subsection{Cadence for recovery and coordinated characterization}

The baseline program follows external discoveries rather than conducting a
blind MIR census. Ground optical surveys submit astrometry to the Minor Planet
Center (MPC), which distributes alerts and preliminary ephemerides. The 3.5ST
scheduler converts the ephemeris and covariance into short exposures that
limit trailing. A visit should contain at least four astrometric measurements
over a time span that resolves the angular motion. A second visit on another
night or at a later geometry extends the orbital arc. CODES determines the
repeat interval from the predicted covariance rather than imposing one fixed interval on every
target.

The 145\,s exposure used in the single-visit sensitivity calculation below
provides a reference depth. The exposure does not define the operational
cadence. Fast and nearby targets require shorter integrations while faint
targets require a longer sequence that preserves measurable motion between
frames. The observing system records the mid-exposure UTC, observatory state,
filter, detector location, and distortion solution for every measurement.

\begin{table}[H]
\centering
\caption{Reference sequence for an externally discovered moving object.
The timing is selected from the orbit covariance and angular rate.}
\begin{tabularx}{\textwidth}{@{}P{35mm}P{48mm}Y@{}}
\toprule
\textbf{\color{cGold}Stage} &
\textbf{\color{cGold}Observation} &
\textbf{\color{cGold}Decision product}\\
\midrule
Ground discovery &
Tracklet, photometry, preliminary orbit, and covariance submitted through
the MPC &
Target priority and a predicted 3.5ST search region\\
3.5ST recovery &
At least four untrailed optical or near-infrared frames with calibrated
timing and distortion &
ICRF positions, angular rate, light-curve points, and an updated orbit\\
3.5ST characterization &
Multiband photometry or spectroscopy at selected rotational phases &
Taxonomy, color, phase behavior, and reflected-light constraints\\
Ground MIR follow-up &
Targeted thermal photometry through an available atmospheric window &
Joint diameter and albedo posterior through Table~\ref{tab:ground-mir-interface}\\
\bottomrule
\end{tabularx}
\label{tab:neo-coordinated-sequence}
\end{table}

The characterized-target yield depends on the external alert rate, the
fraction visible within the 3.5ST field of regard, the Target-of-Opportunity
allocation, and the success rate of ground MIR scheduling. A five-year MIR
discovery yield cannot be assigned to the 3.5ST concept because the mission
does not carry the detector assumed by the former calculation. The proposal
will quote a characterized-target yield only after a joint scheduler applies
the 3.5ST allocation and participating MIR availability to a representative
ground alert stream.

\begin{resultbox}
\textbf{\color{cGold}Cadence conclusion.}
Four measurements provide a practical minimum for one astrometric visit.
The repeat time remains target dependent. The NEO program measures success
through recovered objects, extended orbital arcs, and completed physical
characterization rather than through a 3.5ST MIR discovery count.
\end{resultbox}

\subsection{Single-exposure detection area and ideal warning time}

\begin{figure}[H]
\centering
\includegraphics[width=\textwidth]{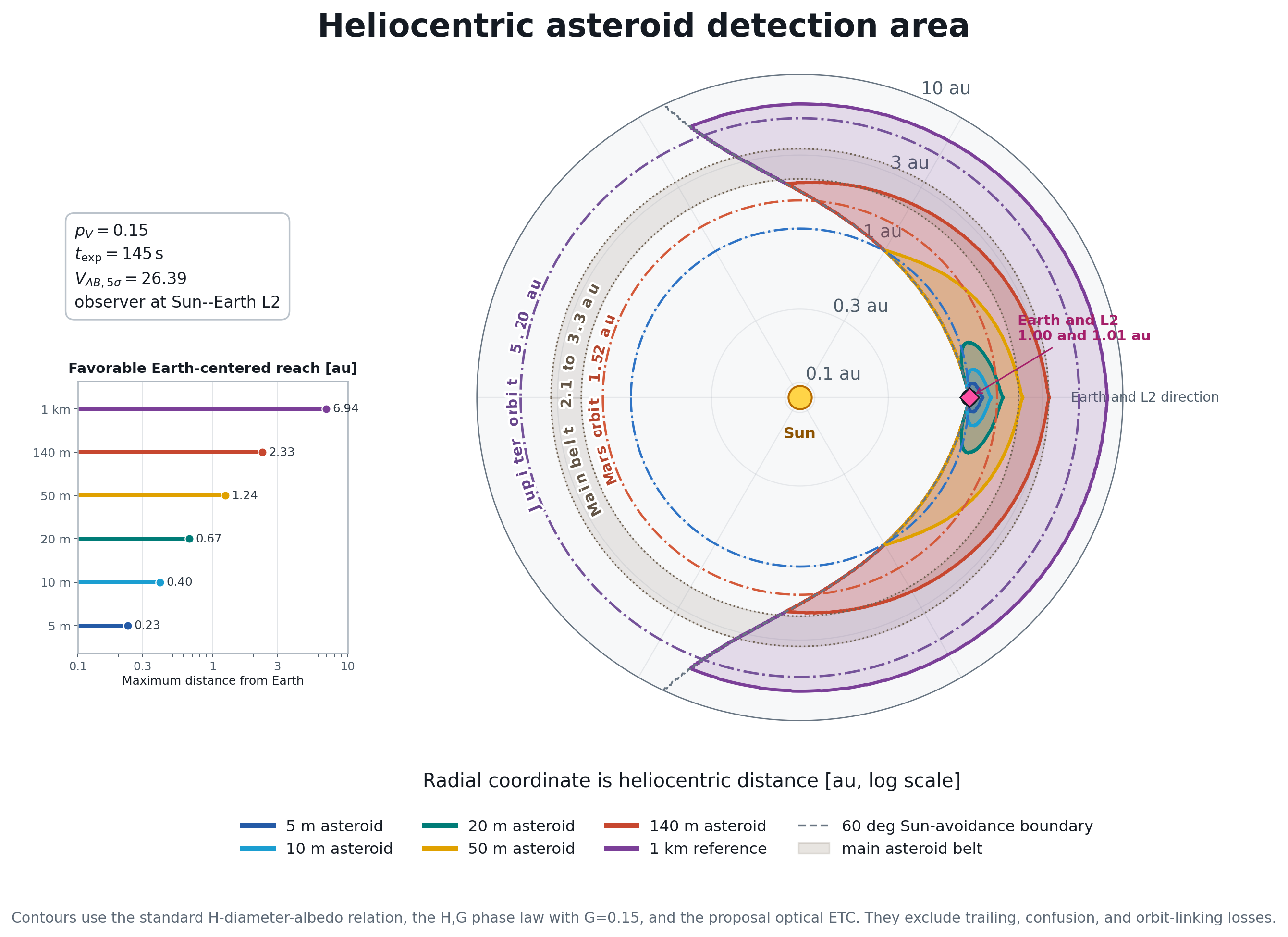}
\caption{Sun-centered single-exposure detection area for reflected light.
The radial coordinate is heliocentric distance on a logarithmic scale. The
orbit guides mark Earth at $1.00$\,au, Mars at $1.52$\,au, the main asteroid
belt from $2.1$ to $3.3$\,au, and Jupiter at $5.20$\,au. The colored contours
use a 145\,s Johnson $V$ exposure, $p_V=0.15$, and the $H,G$ phase law. The
$140$\,m contour reaches the near side of the main belt under favorable
illumination. The 1\,km contour extends beyond Jupiter's orbit near
opposition. The inset reports the favorable maximum Earth-centered distance
for each diameter. The contours are instantaneous flux limits and omit
trailing, confusion, cadence, linking losses, and orbit uncertainty.}
\label{fig:neo-single-visit-reach}
\end{figure}

Figure~\ref{fig:neo-single-visit-reach} converts the optical ETC limit into
a heliocentric detection area. The calculation uses a 145\,s Johnson $V$
exposure with a $5\sigma$ limit of $V_{\rm AB}=26.39$. The geometric albedo
is $p_V=0.15$. The standard diameter relation in
Equation~\ref{eq:h-diameter} and the $H,G$ phase law with $G=0.15$ determine
the reflected flux \cite{Bowell1989}. The observer is placed at Sun--Earth
L2, 0.01\,au beyond Earth.

The Sun-centered panel marks the orbital radii of Mars and Jupiter together
with the main asteroid belt. The $140$\,m contour intersects the
$2.1$--$3.3$\,au main belt where the phase angle and observer distance are
favorable. Asteroids near 140\,m can therefore exceed the single-exposure
threshold on the accessible near side of the belt. The 1\,km contour crosses
the $5.20$\,au orbit of Jupiter and extends beyond it
near opposition. The inset gives the corresponding maximum distance from
Earth for each diameter. The Earth-centered distance sets the ideal
warning-time scale calculated below.

The direction of greatest optical reach does not represent the arrival
direction of every impactor. Reflected-light surveys reach their greatest
distance near opposition and therefore impose a strong opposition selection
on the resulting catalog. Real impactors approach from a broad range of solar
elongations. The 17--20\,m Chelyabinsk impactor approached within
$15^\circ$ of the Sun and could not be detected by conventional survey
telescopes in that direction \cite{JPLChelyabinsk2013}. The solar keep-out
zone therefore prevents a nonzero guaranteed warning time for every
possible impactor.

The maximum Earth-centered detection distance $d_{\oplus,\max}$ sets the
ideal geometric lead time
\begin{equation}
 t_{\rm lead}
 \simeq
 \frac{d_{\oplus,\max}}{v_{\rm rel}}
 =
 57.7\,{\rm d}
 \left(\frac{d_{\oplus,\max}}{1\,{\rm au}}\right)
 \left(\frac{30\,{\rm km\,s^{-1}}}{v_{\rm rel}}\right).
 \label{eq:ideal-warning-time}
\end{equation}
Table~\ref{tab:ideal-warning-time} evaluates Equation~
\ref{eq:ideal-warning-time} for representative approach speeds. The
20\,m and 50\,m rows correspond approximately to the Chelyabinsk and
Tunguska size scales \cite{NASAEmergency2021}. The 140\,m row marks the
NASA population-survey threshold \cite{NASASDT2017}.

\begin{table}[H]
\centering
\caption{Ideal warning time in favorable opposition geometry. The two speed
columns bracket a reference approach-speed range. Values assume continuous
coverage, immediate processing, and successful linkage.}
\begin{tabularx}{\textwidth}{@{}P{20mm}P{45mm}P{32mm}Y@{}}
\toprule
\textbf{\color{cGold}Diameter} &
\textbf{\color{cGold}Reference role} &
\textbf{\color{cGold}$d_{\oplus,\max}$} &
\textbf{\color{cGold}Lead time at 20 and 30 km s$^{-1}$}\\
\midrule
5\,m & Small-object sensitivity & 0.233\,au & 20.2 and 13.4 days\\
10\,m & Warning threshold scale & 0.404\,au & 35.0 and 23.3 days\\
20\,m & Chelyabinsk scale & 0.671\,au & 58.1 and 38.7 days\\
50\,m & Tunguska scale & 1.239\,au & 107.3 and 71.5 days\\
140\,m & NASA census threshold & 2.329\,au & 201.6 and 134.4 days\\
1\,km & Large-object reference & 6.945\,au & 601.2 and 400.8 days\\
\bottomrule
\end{tabularx}
\label{tab:ideal-warning-time}
\end{table}

The tabulated times are upper bounds for one favorable trajectory family.
Reliable alerts require a ground discovery tracklet, successful 3.5ST
recovery, cross-night linkage, and an orbit solution. Continuous all-sky
coverage is not provided by the baseline design. Ground-based MIR
measurements remain important after recovery because reflected light
constrains $p_VD^2$ rather than diameter alone.

\begin{resultbox}
\textbf{\color{cGold}Warning-time interpretation.}
The baseline ETC predicts that a favorable 20\,m object can cross the
detection threshold tens of days before impact within the idealized speed
range. The upper bound does not define a guaranteed warning requirement.
Solar-direction approaches can remain hidden until impact. A defensible
alert claim requires deterministic sky coverage,
real-time processing, linkage efficiency, and a quantified solar blind
zone.
\end{resultbox}

\section{Orbit Determination and Non-Gravitational Forces}

The orbit state $\mathbf{x}=(\mathbf{r},\mathbf{v})$ is propagated with
\begin{equation}
 \ddot{\mathbf r}
 = \sum_i GM_i
 \frac{\mathbf r_i-\mathbf r}{|\mathbf r_i-\mathbf r|^3}
 +\mathbf a_{\rm zonal}
 +\mathbf a_{\rm 1PN}
 +\mathbf a_{\rm SRP}
 +\mathbf a_{\rm PR}
 +\mathbf a_{\rm SW}
 +\mathbf a_{\rm NG}.
 \label{eq:force-model}
\end{equation}
\textbf{Left-hand side.}
The vector $\mathbf r(t)$ is the barycentric position of the target in the
J2000 International Celestial Reference Frame. Its first derivative is the
barycentric velocity $\mathbf v=\dot{\mathbf r}$. The second derivative
$\ddot{\mathbf r}$ is the total acceleration rather than a force. CODES treats
the target as a massless test particle. The target therefore does not perturb
the planetary ephemeris or the motion of another body. Time is measured on
the TDB-compatible ephemeris scale. Positions and accelerations are evaluated
in km and km\,s$^{-2}$ within the numerical force routine.

\textbf{Newtonian gravity.}
The index $i$ runs over the Sun, Mercury barycenter, Venus barycenter, Earth,
Moon, Mars barycenter, Jupiter barycenter, Saturn barycenter, Uranus
barycenter, Neptune barycenter, and Pluto barycenter. The default calculation
also includes the 16 massive main-belt perturbers in the JPL SB441-N16
ephemeris. The vector $\mathbf r_i(t)$ is the barycentric ephemeris position
of body $i$. The displacement $\mathbf r_i-\mathbf r$ points from the target
toward the perturber and its cubic norm supplies the inverse-square
acceleration with the correct direction. The coefficient $GM_i\equiv\mu_i$
is the measured gravitational parameter of the perturber. The implementation
reads $\mu_i$ directly from the JPL kernel or its associated constant table.
The implementation does not multiply a separate value of $G$ by a tabulated mass. The direct
use of $\mu_i$ follows the precision structure of modern solar-system
ephemerides.

\textbf{Planetary zonal gravity.}
The term $\mathbf a_{\rm zonal}$ contains the axisymmetric $J_2$, $J_4$, and
$J_6$ corrections for bodies with measured coefficients. CODES evaluates the
Legendre expansion in the time-dependent IAU north-pole direction supplied
by the NAIF planetary constants kernel. The zonal acceleration becomes most
important during a close planetary passage. DE440s supplies a planet-system barycenter
for Mars and the outer planets. A precision trajectory within a satellite
system must also load the corresponding planet-center satellite SPK.

\textbf{Relativistic acceleration.}
The term $\mathbf a_{\rm 1PN}$ is the first post-Newtonian correction. CODES
now evaluates the massless-target Einstein--Infeld--Hoffmann (EIH) expression
using every active DE440s/SB441-N16 source. The correction includes source
potentials, velocities, and Newtonian source accelerations. The full
expression reduces exactly to the solar Schwarzschild test-particle
expression when the Sun is the only stationary source
\cite{Newhall1983,Tamayo2020}. The source trajectories remain the prescribed
JPL ephemerides rather than being re-integrated by CODES.

\textbf{Solar radiation terms.}
The outward term $\mathbf a_{\rm SRP}$ denotes direct solar-radiation
pressure. Its amplitude is proportional to the solar radiation pressure at
1\,au, the radiation coefficient $C_R$, the area-to-mass ratio, and the
inverse square of heliocentric distance. The term $\mathbf a_{\rm PR}$ is the
velocity-dependent Poynting--Robertson component of the same radiation field.
The Poynting--Robertson acceleration removes orbital energy through radial
and transverse drag. The term $\mathbf a_{\rm SW}$ transfers momentum from a
radial proton wind after subtracting the target velocity from the wind
velocity. The default density is 5\,cm$^{-3}$ at 1\,au and the default radial
speed is 400\,km\,s$^{-1}$. Both quantities and the non-proton momentum factor
are configurable. The model describes a stationary radial wind. The model does not
predict transient coronal structures. The three terms vanish in the local
calculation when no credible area-to-mass ratio is supplied.

\textbf{Empirical non-gravitational acceleration.}
The term $\mathbf a_{\rm NG}$ represents accelerations constrained by
astrometry but not predicted reliably from the known shape and surface state.
CODES resolves the term into $A_1$, $A_2$, and $A_3$ along the
heliocentric radial, transverse, and orbit-normal unit vectors. The adopted
amplitude scales as the inverse square of heliocentric distance. The
transverse coefficient $A_2$ provides a compact Yarkovsky sensitivity model
for an asteroid. The radial and normal coefficients also support diagnostic
fits for asymmetric cometary outgassing. The comet option replaces the
inverse-square amplitude by the normalized Marsden water-ice law
\cite{Marsden1973}. A specified time lag is applied to the osculating
heliocentric state with a
universal-variable Kepler shift before the sublimation amplitude is
evaluated. A scientific orbit solution must fit the coefficients, lag, and
covariance to astrometry. Nonzero trial values in CODES measure dynamical
sensitivity and do not constitute a detection of a non-gravitational force.

The solar-only limit, retained behind the
\code{--solar-1pn-only} diagnostic switch, is
\begin{equation}
 \mathbf a_{\rm 1PN}
 =\frac{GM_\odot}{c^2r^3}
 \left[
 \left(\frac{4GM_\odot}{r}-v^2\right)\mathbf r
 +4(\mathbf r\cdot\mathbf v)\mathbf v
 \right].
\end{equation}
The implemented multi-body correction for target $i$ is
\begin{equation}
\begin{aligned}
\mathbf a_i^{\rm 1PN}={}&
\sum_{j\ne i}\frac{\mu_j\mathbf r_{ij}}{r_{ij}^{3}c^2}
\left[
4\sum_{k\ne i}\frac{\mu_k}{r_{ik}}
+\sum_{k\ne j}\frac{\mu_k}{r_{jk}}
-v_i^2-2v_j^2+4\mathbf v_i\!\cdot\!\mathbf v_j \right.\\
&\left.
+\frac{3(\mathbf r_{ij}\!\cdot\!\mathbf v_j)^2}{2r_{ij}^2}
+\frac{\mathbf r_{ij}\!\cdot\!\mathbf a_j^{N}}{2}
\right]\\
&+\sum_{j\ne i}\frac{\mu_j}{c^2}
\left[
\frac{\mathbf r_{ij}\!\cdot(4\mathbf v_i-3\mathbf v_j)}
{r_{ij}^{3}}(\mathbf v_i-\mathbf v_j)
+\frac{7\mathbf a_j^{N}}{2r_{ij}}
\right].
\end{aligned}
\end{equation}
Here $\mu_j=GM_j$, $\mathbf r_{ij}=\mathbf r_i-\mathbf r_j$,
$r_{ij}=|\mathbf r_{ij}|$, and $\mathbf v_i$ and $\mathbf v_j$ are
barycentric velocities. The Newtonian source acceleration
$\mathbf a_j^{N}=\sum_{k\ne j}\mu_k(\mathbf r_k-\mathbf r_j)/r_{jk}^3$
contains no post-Newtonian term because such a correction inside this factor would
enter only at order $c^{-4}$. The two sums include the Sun, planetary
barycenters, Earth, Moon, and the enabled large asteroids. The Python and
real128 Fortran implementations agree over a two-day Apophis cross-check to
$4.49\times10^{-8}$\,km in position and
$1.12\times10^{-8}$\,mm\,s$^{-1}$ in velocity.

Direct radiation pressure is proportional to area-to-mass ratio. The
Poynting--Robertson term uses the complete first-order photon momentum
equation \cite{Gustafson1994}. The solar-wind term uses the relative velocity
of a configurable radial proton wind. The terms matter most for dust,
meter-scale material, and high area-to-mass artificial objects. For ordinary
NEOs fitted thermal recoil often dominates the non-gravitational orbit
correction.

\begin{figure}[H]
\centering
\includegraphics[width=0.94\textwidth]{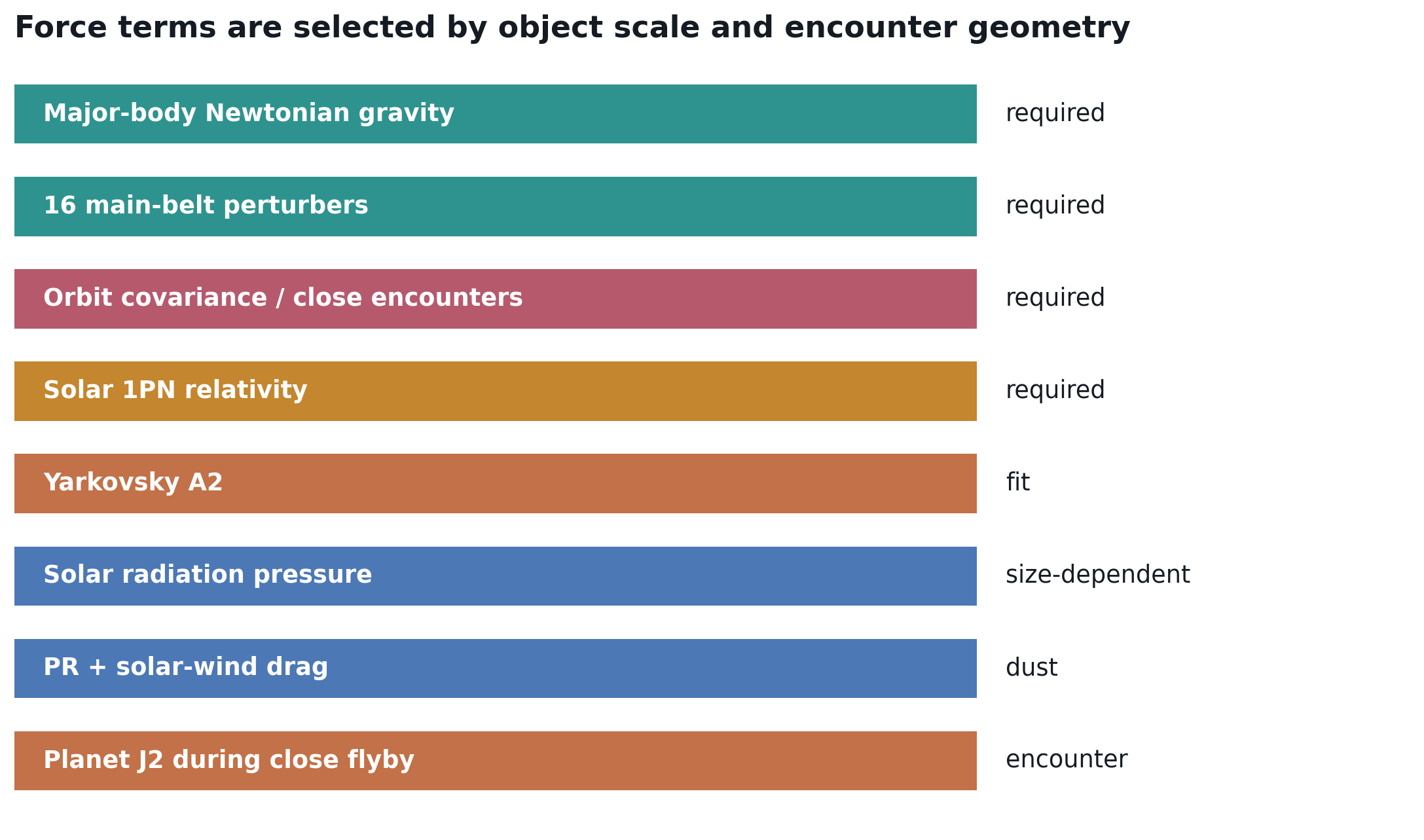}
\caption{Force-model priorities depend on object scale and encounter
geometry. ``Required'' refers to a scientific orbit solution, while
size-dependent terms are enabled only when physical parameters support them.}
\end{figure}

\subsection{Yarkovsky and orbit uncertainty}

The transverse parameter $A_2$ is a compact approximation to thermal recoil
when the acceleration scales approximately as $r^{-2}$. The Bennu orbit
solution measured a semimajor-axis drift near $-19\times10^{-4}$\,au\,Myr$^{-1}$
\cite{Chesley2014}. Interpretation requires diameter, density, spin, thermal
inertia, roughness, and shape. Over a sufficiently long interval YORP can
rotate the spin axis or modify the rotation rate and thereby shift the
magnitude and direction of the thermal recoil.

A 100-year nominal integration does not imply a 100-year hazard prediction.
Close encounters map a small initial covariance into a non-Gaussian future
distribution. A hazard result must propagate virtual asteroids or an
equivalent uncertainty representation and report impact probability. JPL
Sentry-II was designed for this nonlinear problem \cite{Roa2021}.

CODES now requests \code{cov=mat} and \code{full-prec=true} from JPL SBDB,
reads the complete covariance at its own solution epoch, and draws a
reproducible correlated ensemble. The sampled dimensions include the six
heliocentric elements and fitted $A_1$, $A_2$, $A_3$, or time lag when those
parameters occur in the JPL matrix. Every draw is transformed to a
barycentric J2000 state and propagated independently through the same force
model. For JPL Apophis solution 220 the nominal element-to-state conversion
agrees with the independently requested Horizons state at the covariance
epoch to 0.714\,m and $1.43\times10^{-4}$\,mm\,s$^{-1}$. The direct
finite-sample impact fraction is a screening diagnostic, not a replacement
for Sentry-II rare-event searches. Propagating that nominal local state to the
2029 encounter gives 38,042.1\,km versus the CNEOS value 38,011.5\,km, with
local-minus-JPL residuals of $+30.6$\,km and $-228.8$\,s. The complete
uncertainty ensemble therefore propagates as intended. The
residual prevents presenting CODES as a replacement for the registered JPL
force model.

\section{CODES Orbit Dynamics and Ephemeris System}

The Close-approach Orbit Dynamics and Ephemeris System
(CODES)\footnote{\url{https://github.com/kjhan0606/CODES.git}} implements the
accompanying NEO calculations and validation tests. The Python package retains
\code{neo\_orbit\_calculator} for command-line compatibility. The integrated
desktop interface organizes the commands into NEO dynamics, comet-orbit
evolution, apparent sky positions, and validation. The NEO dynamics tab keeps
the authoritative JPL product separate from the CODES sensitivity calculation.

\begin{resultbox}
\textbf{\color{cGold}Authoritative SPK mode.}
The Python interface requests a time-continuous small-body SPK from NASA/JPL
Horizons for the selected interval. A 2026--2126 SPK is the preferred
reproducibility product. SPK segments store piecewise Chebyshev coefficients.
The polynomial degree is not an orbit-integrator order.
\end{resultbox}

\begin{resultbox}
\textbf{\color{cTeal}Local sensitivity mode.}
A Fortran backend calls CSPICE directly at every force evaluation. The
integrated state, force sums, and adaptive Runge--Kutta arithmetic use
\code{selected\_real\_kind(33,4931)}. GNU Fortran provides 33 decimal digits
on the current platform. SPICE input and output remain IEEE-754 binary64,
which is the precision boundary of the JPL SPK interface.
\end{resultbox}

The phrase ``NASA 100th-order integration coefficients'' does not correspond
to the JPL product. DE440 and DE441 are fitted numerical ephemerides delivered
through SPK Chebyshev segments \cite{Park2021}. Increasing a formal local
integrator order above 100 would not remove initial-state covariance,
close-encounter sensitivity, uncertain thermal recoil, or omitted perturbers.

\subsection{Included gravitational bodies}
\label{sec:calculator-bodies}

The baseline gravity model includes 27 sources. Planet-system barycenters
replace individual planets and satellites where the loaded ephemeris provides
only their combined state.

\noindent
\begin{table}
\captionsetup{justification=raggedright, singlelinecheck=false}
\caption{Bodies included in the CODES gravity model, grouped by
category.}
\label{tab:c6-gravity-bodies}
\centering
\begin{tabularx}{\textwidth}{@{}P{43mm}Y@{}}
\toprule
\color{cGold}\textbf{Group} & \color{cGold}\textbf{Bodies}\\
\midrule
Major bodies &
Sun, Mercury barycenter, Venus barycenter, Earth, Moon, Mars barycenter,
Jupiter barycenter, Saturn barycenter, Uranus barycenter, Neptune barycenter,
and Pluto barycenter\\
SB441-N16 &
Ceres, Vesta, Pallas, Hygiea, Davida, Interamnia, Europa, Sylvia, Eunomia,
Juno, Psyche, Camilla, Thisbe, Iris, Euphrosyne, and Cybele\\
\bottomrule
\end{tabularx}
\end{table}

SB441-N16 is not defined by a universal mass threshold. The designation
denotes the JPL set of the 16 most massive main-belt perturbers. Cybele is the
least massive object
in this set with $GM=0.938105756$\,km$^3$\,s$^{-2}$. Conversion with a
separately adopted $G$ gives a reference mass near
$1.41\times10^{19}$\,kg although the integrator uses $GM$ directly.
The SB441 memo also provides a 373-body ephemeris for specialized studies
\cite{Farnocchia2021SB441}.

\subsection{Implemented and deferred force terms}

\begin{table}
\caption{Force terms implemented in the CODES integrator, and the
physical regime in which each term is enabled or deferred.}
\label{tab:c7-force-terms}
\centering
\begin{tabular}{@{}M{43mm}C{37mm}M{85mm}@{}}
\toprule
\color{cGold}\textbf{Term} &
\color{cGold}\textbf{Status} &
\color{cGold}\textbf{Interpretation}\\
\midrule
Major-body gravity and SB441-N16 &
\color{cTeal}\textbf{Implemented} &
DE440s and SB441-N16 positions with JPL $GM$ values. Exercised by the
ten-object close-approach suite\\
\cmidrule(lr){1-3}
Solar Schwarzschild 1PN &
\color{cTeal}\textbf{Implemented} &
Retained as the \code{--solar-1pn-only} limiting-case diagnostic\\
\cmidrule(lr){1-3}
Solar radiation pressure &
\color{cTeal}\textbf{Implemented} &
Enabled when the user supplies an area-to-mass ratio and radiation
coefficient\\
\cmidrule(lr){1-3}
Poynting--Robertson and solar-wind drag &
\color{cTeal}\textbf{Implemented} &
First-order photon momentum equation and configurable radial proton-wind
momentum transfer\\
\cmidrule(lr){1-3}
$A_1,A_2,A_3$ &
\color{cTeal}\textbf{Implemented} &
Configurable radial, transverse, and normal accelerations with $r^{-2}$
scaling\\
\cmidrule(lr){1-3}
Full orbit covariance and virtual asteroids &
\color{cTeal}\textbf{Implemented} &
Full JPL solution-epoch matrix, correlated seeded draws, fitted
non-gravitational parameters, and independent clone propagation. Direct
Monte Carlo remains a screening result\\
\cmidrule(lr){1-3}
Planetary $J_2$ and higher harmonics &
\color{cTeal}\textbf{Implemented} &
Axisymmetric $J_2$, $J_4$, and $J_6$ in the time-dependent IAU pole
direction\\
\cmidrule(lr){1-3}
Full multi-body 1PN formulation &
\color{cTeal}\textbf{Implemented} &
Massless-target EIH correction from every active source, with separate
Python and real128 Fortran implementations\\
\cmidrule(lr){1-3}
Comet outgassing law and time lag &
\color{cTeal}\textbf{Implemented} &
Normalized Marsden water-ice law with configurable coefficients and a
universal-variable Kepler time shift\\
\cmidrule(lr){1-3}
Lorentz force and radiation shadowing &
\color{cOrange}\textbf{Not implemented} &
Required only for charged grains and eclipse geometries\\
\bottomrule
\end{tabular}
\end{table}

The status refers to the CODES integrator. An authoritative Horizons SPK may
contain additional registered force terms in the JPL orbit solution. An
imported SPK can therefore include accelerations that the CODES force model
does not reproduce independently.

\subsection{Reproducibility test}

Every CODES release should record the JPL orbit-solution identifier, solution date,
DE and SB kernel names, kernel checksums, TDB epochs, compiler version,
floating-point kind, tolerances, force switches, and source revision. The
CODES state must be compared with Horizons on an independent epoch grid.
Residuals should be reported before and after close encounters. A residual
threshold must be set by the science case, not by the number of printed
digits.

The real128 Fortran step is selected by both the embedded
Dormand--Prince local-error estimate and a physical close-passage limiter.
For every active source, the internal step is no larger than 5\% of the
gravity timescale $\sqrt{r^3/GM}$ or the relative crossing time
$r/|\Delta\mathbf v|$. High acceleration or high relative speed therefore
forces smaller steps independently of the requested output cadence.

\subsection{Ten historical close-approach tests}
\label{sec:historical-ca-tests}

The regression suite uses ten real NEOs observed between 2019 and 2024.
For each object the calculator downloads all available optical ADES
astrometry from the Minor Planet Center and retains the observatory codes,
UTC epochs, reported coordinates, and MPC discovery flag
\cite{MPCObservations}. The observing arc is therefore traceable to the
measurements used by the international small-body community.

The four-day experiment isolates \emph{trajectory propagation}. The current JPL Horizons
solution supplies the barycentric initial state two days before the official
closest-approach epoch. The real128 Fortran model then propagates for four
days with DE440s, SB441-N16, and solar 1PN enabled. A cubic interpolation of
the propagated state locates the Earth-centered distance minimum between
output epochs. The official epoch, miss distance, and relative speed come
from the JPL CNEOS Close-Approach Data API \cite{JPLCAD}. Heliocentric
ecliptic osculating elements are compared with Horizons one day before the
encounter, before the strongest terrestrial deflection.

\begin{resultbox}
\textbf{\color{cGold}Source convention.}
MPC supplies the observed astrometric history, station count, and discovery
epoch. NASA/JPL supplies the official current initial state, osculating
elements, closest-approach epoch, and geocentric distance. Every reported
$\Delta$ is the \textbf{CODES result minus the NASA/JPL official result}.
The source snapshot used here was retrieved on 2026-07-26 UTC.
\end{resultbox}

\begin{table}[H]
\centering
\footnotesize
\caption{Historical close-approach reproduction. $N_{\rm obs}/N_{\rm stn}$ and lead time are measured from Minor Planet Center (MPC) ADES observations \cite{MPCObservations}; the official epoch and geocentric distance are NASA/JPL CNEOS CAD values \cite{JPLCAD}. Residuals are CODES minus JPL. Positive lead time means discovery before closest approach.}
\label{tab:neo-ca-validation}
\begin{tabularx}{\textwidth}{@{}l r r r r r r@{}}
\toprule
\color{cGold}\textbf{Object} & \color{cGold}\textbf{$N_{\rm obs}/N_{\rm stn}$} & \color{cGold}\textbf{Lead [h]} & \color{cGold}\textbf{JPL CA [UTC]} & \color{cGold}\textbf{$d_{\rm JPL}$ [km]} & \color{cGold}\textbf{$\Delta t$ [s]} & \color{cGold}\textbf{$\Delta d$ [km]}\\
\midrule
2019 OK & 82/9 & +24.2 & 2019-07-25 01:21 & 71355.0 & -0.001 & +0.000\\
2019 UN13 & 18/4 & +8.0 & 2019-10-31 14:44 & 12613.4 & -0.000 & -0.003\\
2020 QG & 35/9 & -6.3 & 2020-08-16 04:07 & 9316.9 & +0.002 & -0.001\\
2020 VT4 & 34/9 & -15.4 & 2020-11-13 17:19 & 6745.5 & +0.002 & -0.000\\
2021 GW4 & 40/8 & +103.5 & 2021-04-12 13:00 & 26199.3 & -0.002 & -0.001\\
2021 UA1 & 23/4 & -3.9 & 2021-10-25 03:05 & 9426.7 & -0.002 & -0.002\\
2022 NF & 31/8 & +77.6 & 2022-07-07 13:43 & 88993.9 & -0.113 & -0.000\\
2023 BU & 1784/38 & +120.6 & 2023-01-27 00:27 & 9967.0 & +0.098 & +0.000\\
2024 MK & 843/60 & +306.2 & 2024-06-29 13:47 & 295419.9 & +0.330 & -0.000\\
2024 ON & 912/67 & +1248.7 & 2024-09-17 10:17 & 1000052.6 & -0.007 & -0.002\\
\bottomrule
\end{tabularx}
\end{table}

\begin{table}[H]
\centering
\small
\caption{Osculating-element residuals one day before closest approach. Columns labeled JPL are the current NASA/JPL Horizons heliocentric ecliptic solution \cite{Horizons}; each $\Delta$ is CODES minus JPL at the same TDB epoch.}
\label{tab:neo-element-validation}
\begin{tabularx}{\textwidth}{@{}l r r r r r r@{}}
\toprule
\color{cGold}\textbf{Object} & \color{cGold}\textbf{$a_{\rm JPL}$ [au]} & \color{cGold}\textbf{$\Delta a$ [km]} & \color{cGold}\textbf{$e_{\rm JPL}$} & \color{cGold}\textbf{$10^9\Delta e$} & \color{cGold}\textbf{$i_{\rm JPL}$ [deg]} & \color{cGold}\textbf{$\Delta i$ [mas]}\\
\midrule
2019 OK & 1.9484776 & +0.001 & 0.7621856 & +0.001 & 1.40156 & +0.000\\
2019 UN13 & 0.9884240 & -0.000 & 0.3474246 & +0.002 & 1.49958 & +0.000\\
2020 QG & 1.9630290 & +0.002 & 0.4925217 & +0.004 & 5.50775 & +0.000\\
2020 VT4 & 1.3177933 & -0.001 & 0.2492251 & -0.003 & 13.02271 & +0.000\\
2021 GW4 & 1.5250861 & -0.001 & 0.3596578 & -0.003 & 0.74254 & -0.000\\
2021 UA1 & 0.9827608 & -0.000 & 0.3757717 & +0.002 & 10.21981 & +0.001\\
2022 NF & 2.3967733 & +0.005 & 0.6073175 & +0.006 & 1.29947 & -0.000\\
2023 BU & 0.9854978 & +0.006 & 0.0776724 & +0.005 & 2.76782 & +0.000\\
2024 MK & 2.2477429 & -0.002 & 0.5510890 & -0.003 & 8.48901 & +0.000\\
2024 ON & 2.3881172 & +0.002 & 0.5783376 & +0.002 & 7.74315 & -0.000\\
\bottomrule
\end{tabularx}
\end{table}

\begin{figure}[H]
\centering
\includegraphics[width=\textwidth]{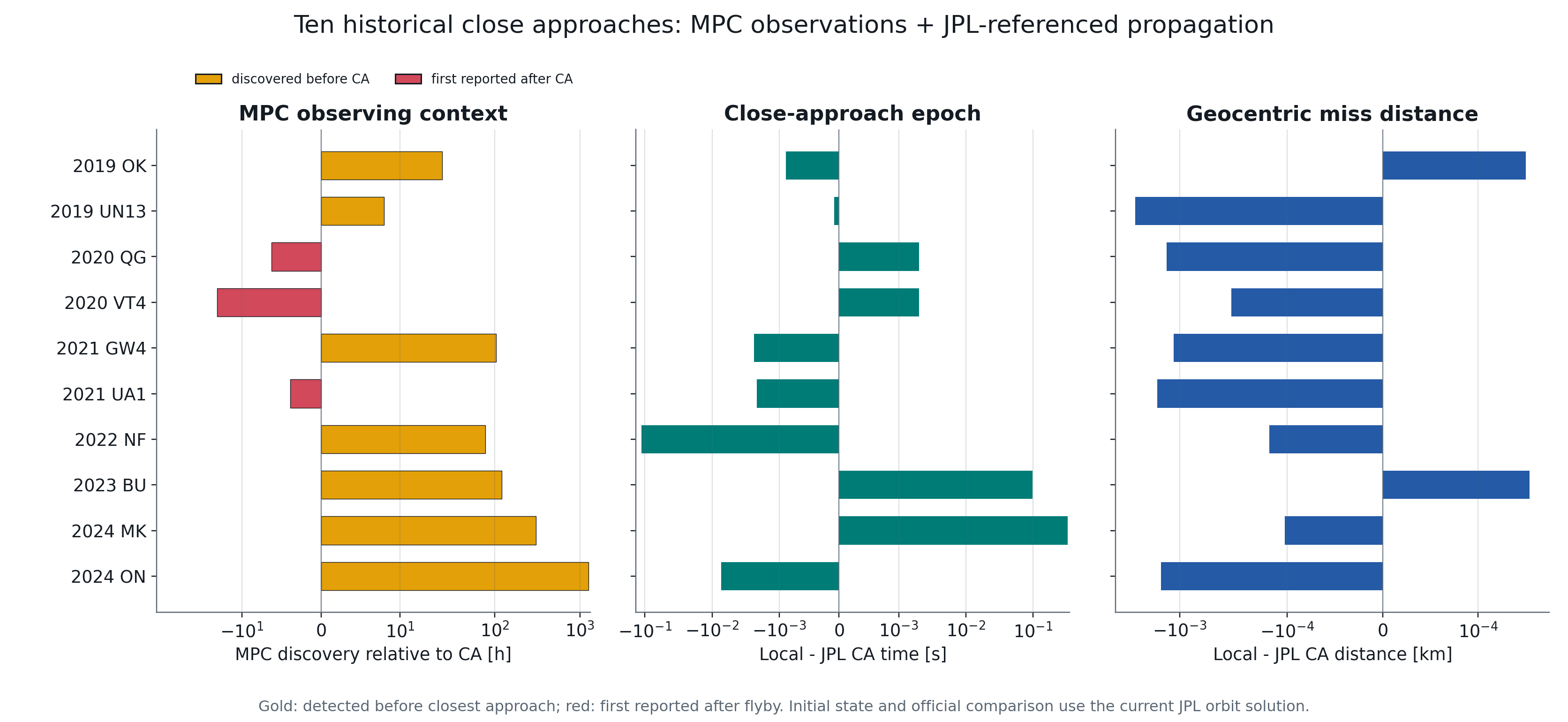}
\caption{Observation context and propagation residuals for ten historical
flybys. Gold bars in the left panel indicate discovery before closest
approach. Red bars indicate that the first discovery observation followed
the flyby. All three horizontal axes use symmetric-log scaling so that zero,
sign, and decade-level differences remain visible. The linear thresholds are
12\,h, $10^{-3}$\,s, and $10^{-4}$\,km from left to right. The center and
right panels compare the CODES force model with the current official JPL
solution. Across the ten cases the maximum absolute
residual is 0.330\,s in closest-approach epoch, 0.00259\,km in miss distance,
and 0.00561\,km in pre-encounter semimajor axis.}
\label{fig:neo-orbit-validation}
\end{figure}

\begin{warningbox}
\textbf{\color{cOrange}Validation boundary.}
The JPL initial state is based on the current orbit solution and can include
post-encounter observations. Tables~\ref{tab:neo-ca-validation} and
\ref{tab:neo-element-validation} therefore certify numerical propagation
against an authoritative trajectory. The tables do not certify an
independent orbit fit from a discovery-night arc. Operational acceptance also
requires observatory-state reconstruction, catalog-bias correction,
weighted batch least squares, a six-dimensional covariance, outlier
rejection, and archived discovery-epoch solutions for a blind backtest.
\end{warningbox}

\subsection{Medium- and long-period comet modes}

CODES provides two distinct comet products. The \code{comet-orbits} mode
assembles the separate NASA/JPL Horizons solution for each identified
apparition and plots the return interval, semimajor axis, and perihelion
distance. The \code{comet-sky} mode accepts specified UTC epochs and computes
the geocentric right ascension, declination, distance, and IAU constellation.
The first product quantifies long-term dynamics. The second supplies the sky
position needed to schedule an observation.

A return of 1P/Halley means a passage through perihelion rather than a return
to Earth. The present JPL \#75 solution uses 8,518 observations over
1835--1994. The long-term propagation test takes only its barycentric state
at 1850 January 2 TDB and the published non-gravitational coefficients
$A_1=4.887055233121\times10^{-10}$ and
$A_2=1.554720290005\times10^{-10}$\,au\,d$^{-2}$. CODES then propagates to
2000 with its real128 Fortran integrator, DE440s major-body states,
SB441-N16 perturbers, full multi-body 1PN, and the standard Marsden water-ice
law \cite{Marsden1973}.
No JPL comparison state is ingested after the initial epoch.

\begin{figure}[H]
\centering
\includegraphics[width=\textwidth]{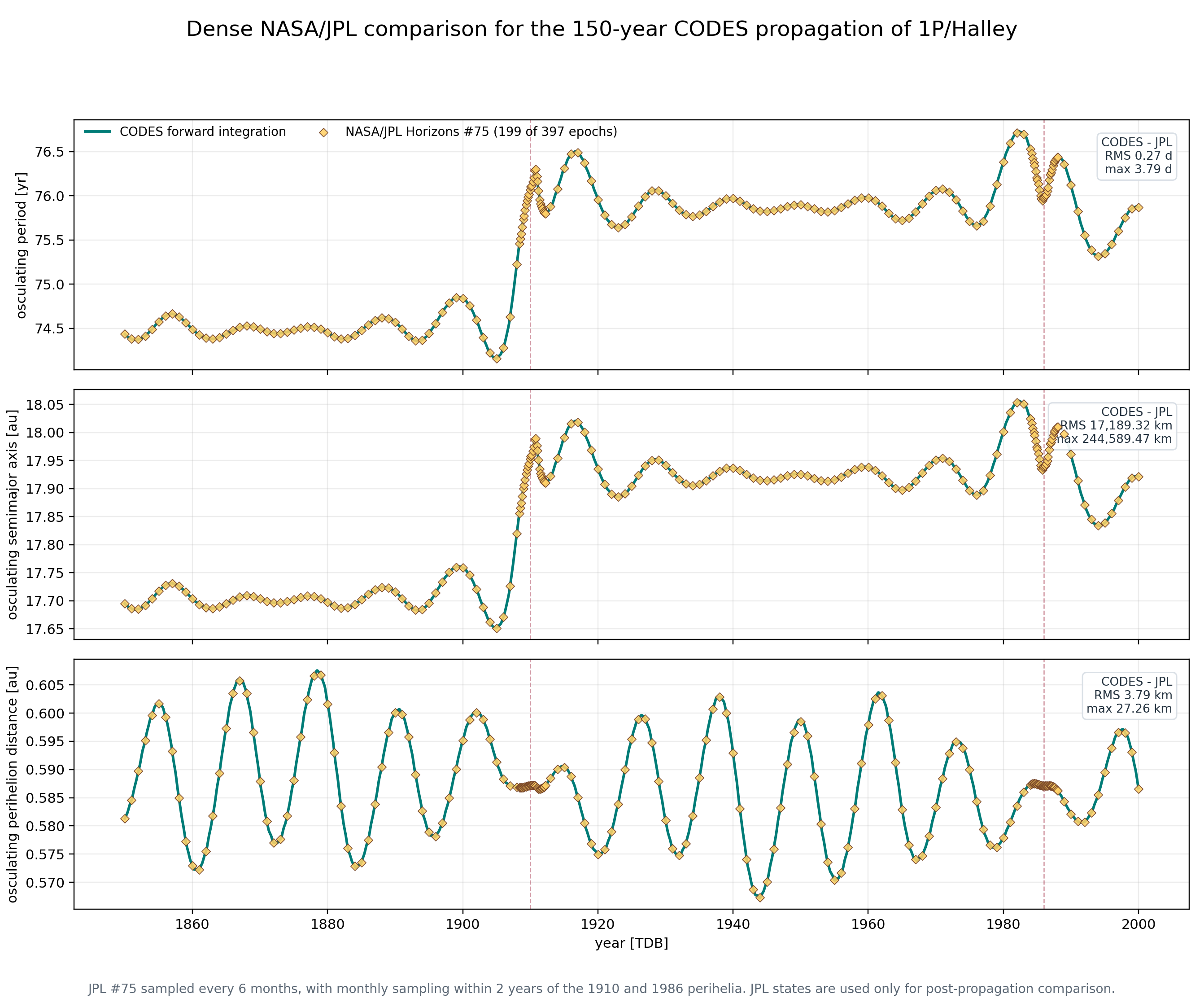}
\caption{Independent 150-year CODES propagation of 1P/Halley. Open circles
are CODES results and gold diamonds are separately requested NASA/JPL
Horizons comparison values in every panel. From top to bottom, the continuous
curves show the osculating period, semimajor axis, and perihelion distance
derived from the CODES state. The CODES minus JPL perihelion-time residuals
are $-12.956$\,s and $+8.835$\,s. The corresponding period residuals are
$-3.840$\,d and $-1.931$\,d. The semimajor-axis residuals are
$-247{,}460$\,km and $-124{,}487$\,km. The perihelion-distance residuals are
$+227.8$\,km and $-15.3$\,km.}
\label{fig:halley-return-history}
\end{figure}

\begin{resultbox}
\textbf{\color{cGold}Validated quantity.}
The comparison tests long-term numerical propagation from one initial state
rather than the reconstruction of a sequence of JPL records. The agreement
shows that CODES reproduces the two physical perihelion passages across
150 years.
Because JPL solution \#75 was fitted with observations extending through
1994, the comparison is a propagation-consistency test rather than a blind
historical orbit determination.
\end{resultbox}

Replacing the solar-only 1PN term by the full multi-body expression shifts
the 1986 perihelion-time residual by approximately $-0.20$\,s and the
semimajor-axis residual by approximately $-3.5$\,km. The remaining
$1.24\times10^5$\,km semimajor-axis difference is therefore not explained by
the former relativistic approximation alone. Covariance propagation also does
not shift the nominal curve. The covariance supplies an
uncertainty ensemble around the same fitted central solution. Adding the
physical adaptive-step limiter
shifts the 150-year trajectory by at most 13.6\,m and the final position by
1.11\,m relative to error control alone, leaving the quoted passage values
unchanged. Resolving the remaining difference requires a controlled comparison
of the fitted comet non-gravitational law, source ephemeris, and JPL solution
metadata.

DE440s does not support a defensible extension of the modern-state integration
through the entire ancient record. CODES therefore retains a separate
historical product for returns before the supported interval. JPL apparition
solutions are compared with the observation-constrained perihelion times of Yeomans and
Kiang \cite{Horizons,YeomansKiang1981}. The records diagnose period evolution
without relabeling an unconstrained modern-state extrapolation as an observed
trajectory.

\begin{table}[H]
\centering
\caption{CODES reconstruction of selected 1P/Halley perihelia. The residual is the calculated time minus the historical observation-based time reported by Yeomans and Kiang \cite{Horizons,YeomansKiang1981}.}
\begin{tabularx}{\textwidth}{@{}P{14mm}P{44mm}P{29mm}P{28mm}Y@{}}
\toprule
\color{cGold}\textbf{Year} &
\color{cGold}\textbf{Horizons perihelion, TDB} &
\color{cGold}\textbf{Return interval} &
\color{cGold}\textbf{$\Delta a$} &
\color{cGold}\textbf{Comparison}\\
\midrule
1986 & 1986-Feb-08.474 & 75.806\,yr & -0.030\,au & NASA/JPL\\
1910 & 1910-Apr-20.178 & 74.422\,yr & -0.031\,au & NASA/JPL\\
1835 & 1835-Nov-16.440 & 76.677\,yr & -0.098\,au & NASA/JPL\\
1759 & 1759-Mar-13.062 & 76.487\,yr & -0.082\,au & NASA/JPL\\
1607 & 1607-Oct-27.541 & 76.143\,yr & -0.070\,au & NASA/JPL\\
1531 & 1531-Aug-26.239 & 75.210\,yr & -0.094\,au & +0.44\,d\\
1145 & 1145-Apr-18.561 & 79.079\,yr & -0.038\,au & -2.69\,d\\
1066 & 1066-Mar-20.934 & 76.537\,yr & +0.332\,au & -2.57\,d\\
989 & 989-Sep-05.688 & 77.134\,yr & -0.047\,au & -3.31\,d\\
912 & 912-Jul-18.674 & 75.385\,yr & +0.080\,au & +9.17\,d\\
837 & 837-Feb-28.270 & -- & -- & -1.83\,d\\
\bottomrule
\end{tabularx}
\label{tab:halley-returns}
\end{table}

\begin{resultbox}
\textbf{\color{cGold}Largest adjacent orbital shift.}
The largest shift between adjacent solutions occurs from 989 to 1066. The
semimajor axis increases by 0.332\,au and the osculating period
increases by 2.133\,yr. The subsequent 1066--1145 return interval is
79.079\,yr. The 1066 historical solution places the minimum separation from
the Jupiter barycenter at 1.614\,au on 1067 March 5 TDB. The separation is not
a close encounter and the complete period increase cannot be assigned to one
Jupiter passage. The fitted orbit includes accumulated planetary and
non-gravitational perturbations. Yeomans and Kiang instead identify the
0.04\,au Earth encounter in 837 as the severe ancient rectification boundary
\cite{YeomansKiang1981}.
\end{resultbox}

\begin{verbatim}
python -m neo_orbit_calculator.cli spk 99942 \
  --start 2026-01-01 --stop 2126-01-01

python -m neo_orbit_calculator.cli propagate 99942 \
  --start 2026-01-01 --stop 2126-01-01 \
  --backend fortran

python -m neo_orbit_calculator.cli comet-orbits 1P \
  --start-year 800 --stop-year 2000

python -m neo_orbit_calculator.cli comet-sky 1P \
  --epochs 2061-07-28 2061-08-15 2061-09-01
\end{verbatim}

\section{DS9 to CODES with OpenOrb}
\label{sec:ds9-codes-openorb}

\begin{figure}[H]
\centering
\includegraphics[width=\textwidth]{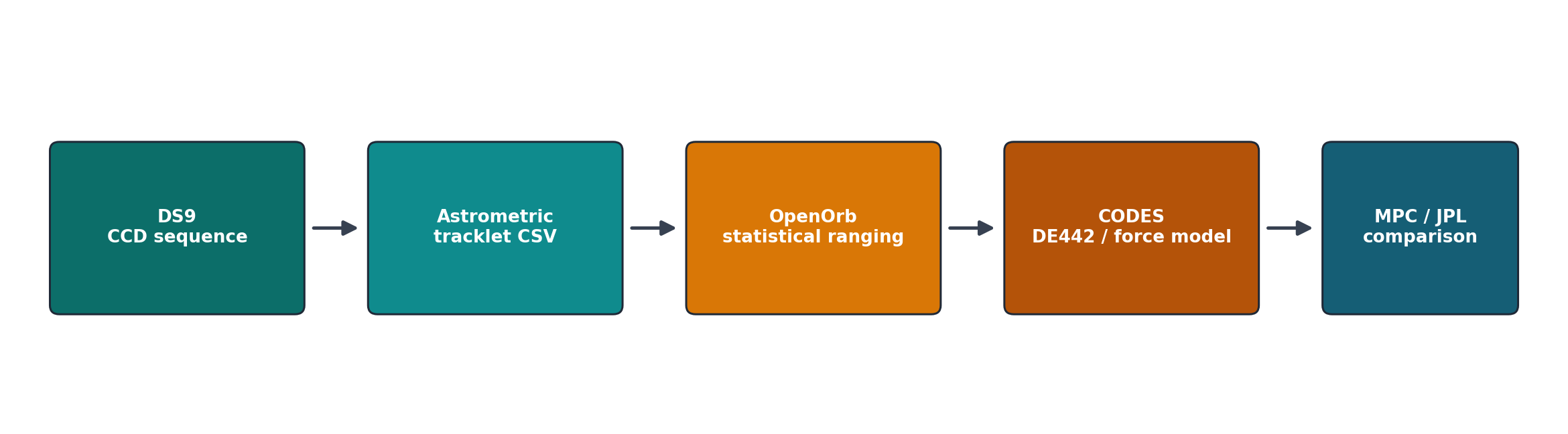}
\caption{Conceptual DS9 to CODES and OpenOrb workflow. A candidate is
reviewed in the time-ordered image sequence before its astrometry is passed
to statistical ranging. The posterior is then propagated by CODES and
compared with an independent official orbit product.}
\label{fig:ds9-codes-openorb-flow}
\end{figure}

Figure~\ref{fig:ds9-codes-openorb-flow} defines the operational sequence from
an image series to an orbit product. DS9 supports visual inspection and
candidate marking. The reviewer exports the accepted positions and times as
an astrometric CSV. CODES passes the CSV to OpenOrb for statistical
ranging, which samples the admissible orbital-element distribution instead
of selecting one distance from a short optical arc \cite{Granvik2009}.
CODES then propagates the samples with the selected DE442 or DE441 kernel
\cite{NAIFDE442,Park2021} and its force model. The final comparison is made
against a separately retrieved NASA/JPL solution. The official solution is
therefore a comparison
target and is not used as an input to the ranging calculation.

\subsection{DS9 image review and astrometric handoff}

The public NOIRLab Astro Data Lab DECam Asteroid Database DR2 provides the
real calibrated DECam sequence used for the bridge demonstration
\cite{DataLabDAD}. The cyan marks and candidate links use the same astrometric
schema that the DS9 NEO menu exports. The implemented menu runs source
extraction and frame-to-frame association before writing an astrometric CSV
and frame-specific DS9 FK5 region files. The observation time, right
ascension, declination, uncertainty, observer code, and output paths are
retained for auditability. Visual review remains mandatory before the
astrometry is passed to CODES and OpenOrb.

\begin{figure}[H]
\centering
\includegraphics[width=\textwidth]{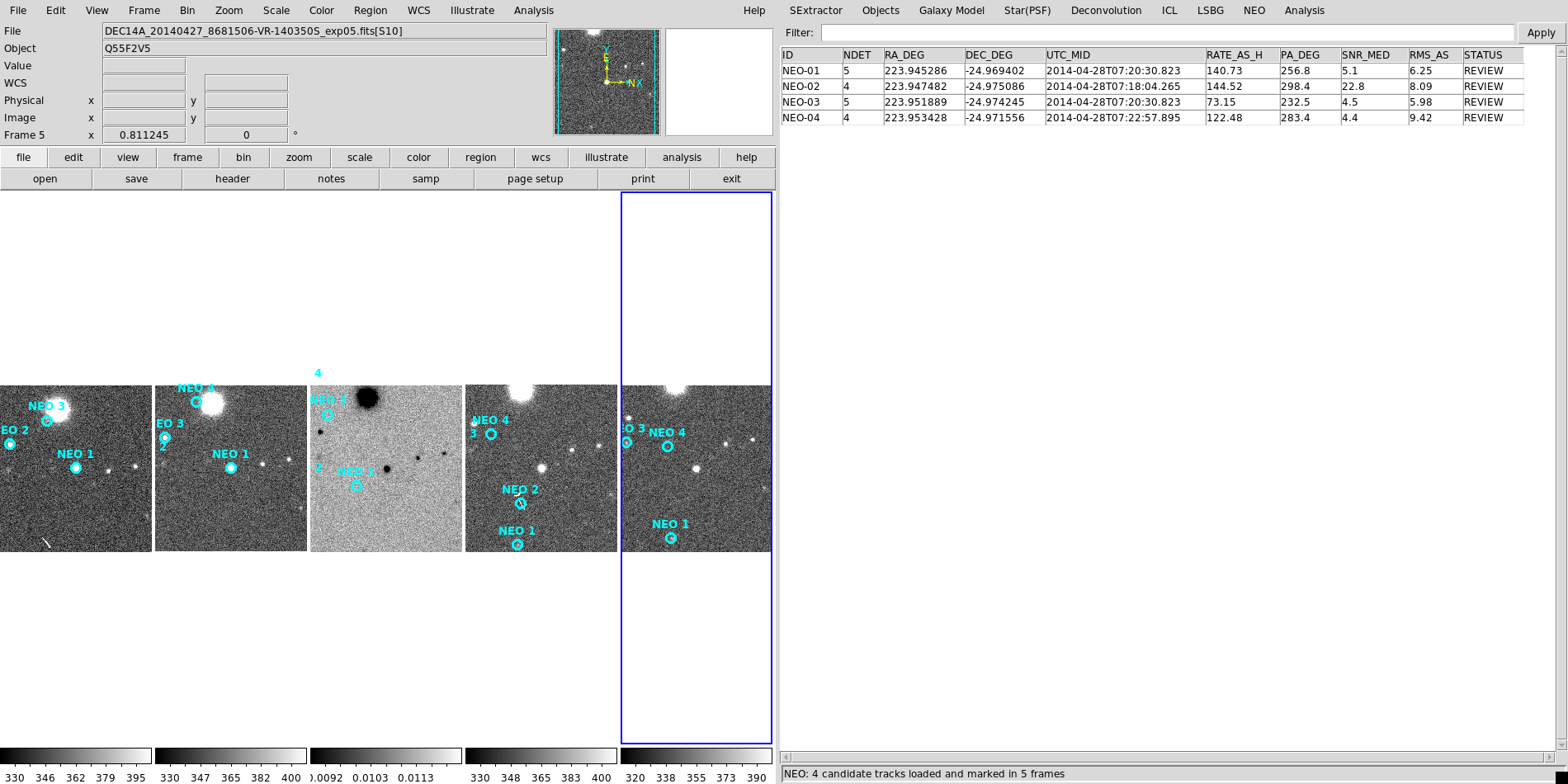}
\caption{Actual OGFinder DS9 runtime screen from five calibrated DECam
exposures in the DECam Asteroid Database DR2. Cyan circles mark four linked
candidate tracklets with each frame showing detections at the corresponding
epoch. The right catalog reports the number of detections, median sky position,
midpoint time, fitted angular rate, position angle, median signal-to-noise
ratio, and astrometric residual for each tracklet. The REVIEW flag identifies
candidates that require visual inspection and orbit confirmation. The status
line records successful association and catalog loading.}
\label{fig:ds9-ogfinder-interface}
\end{figure}

\subsection{OpenOrb ranging and CODES propagation}

\begin{figure}[H]
\centering
\includegraphics[width=\textwidth]{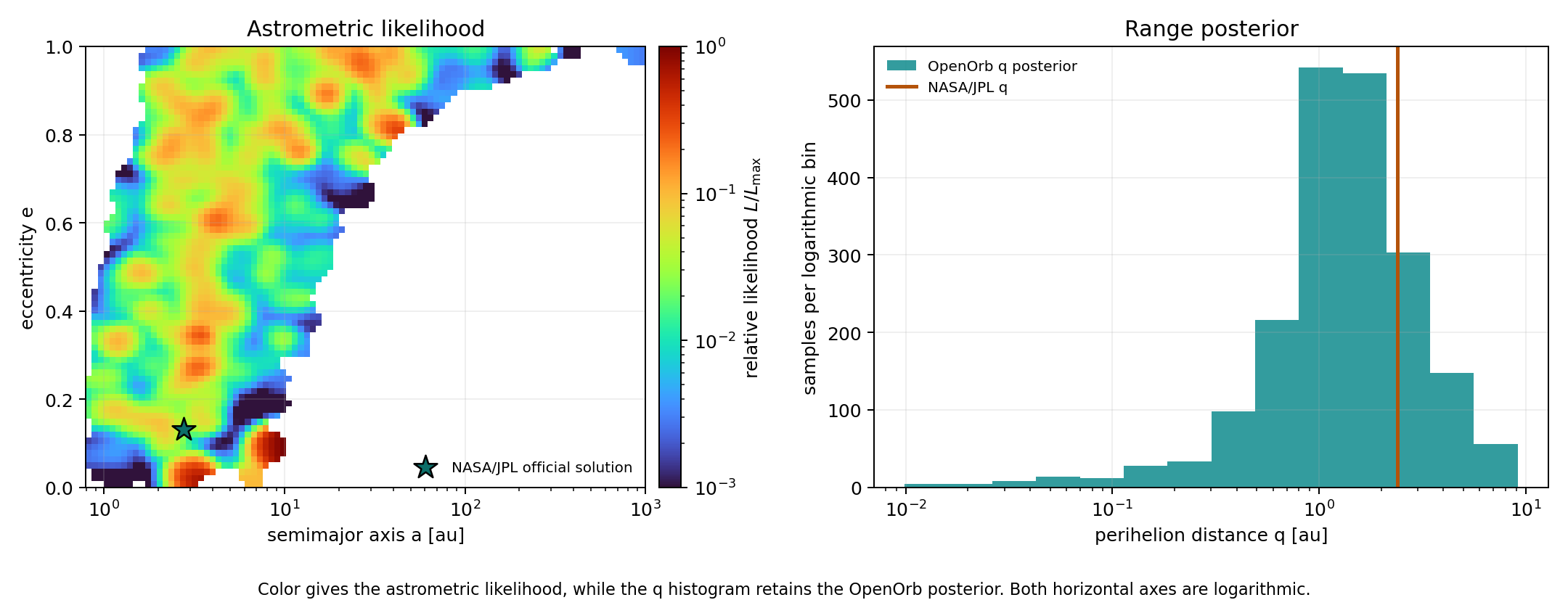}
\caption{OpenOrb statistical ranging output for the DAD demonstration
tracklet. The left panel gives the locally smoothed likelihood field in
$(a,e)$. Color represents $L/L_{\max}$. Regions without sampled support
remain blank. The right panel gives the OpenOrb posterior count in
logarithmic $q$ bins. The horizontal axes are logarithmic. The star and
vertical line mark the separate NASA/JPL solution for the same numbered
object. The official solution does not enter the ranging calculation.}
\label{fig:codes-openorb-calculation}
\end{figure}

OpenOrb performs statistical ranging through its command-line task. CODES
writes an MPC-format observation file, records the exact executable and
configuration file, and preserves the ranging output. The posterior is broad
because five exposures do not constrain range and radial velocity tightly.
The broad distribution does not indicate a failure of the angular astrometry.
Each accepted OpenOrb sample retains its astrometric
$\chi^2$. The additive offset written by OpenOrb cancels in the likelihood
ratio
\[
  \frac{L_i}{L_{\max}} =
  \exp\left[-\frac{\chi_i^2-\chi_{\min}^2}{2}\right].
\]
Figure~\ref{fig:codes-openorb-calculation} shows the sampled likelihood
field in $(a,e)$. Before smoothing the calculation divides the local
likelihood sum by the local sample count. The normalization reduces the
dependence on ranging density and its coordinate Jacobian. The right panel gives the posterior
distribution of $q$. Both horizontal axes use logarithmic scales. The
NASA/JPL point provides an external comparison.
CODES propagates every accepted sample. Later observations can therefore be
tested against the complete distribution rather than one poorly constrained
preliminary orbit.

\subsection{Verification sequence and claim boundary}

\begin{figure}[H]
\centering
\includegraphics[width=\textwidth]{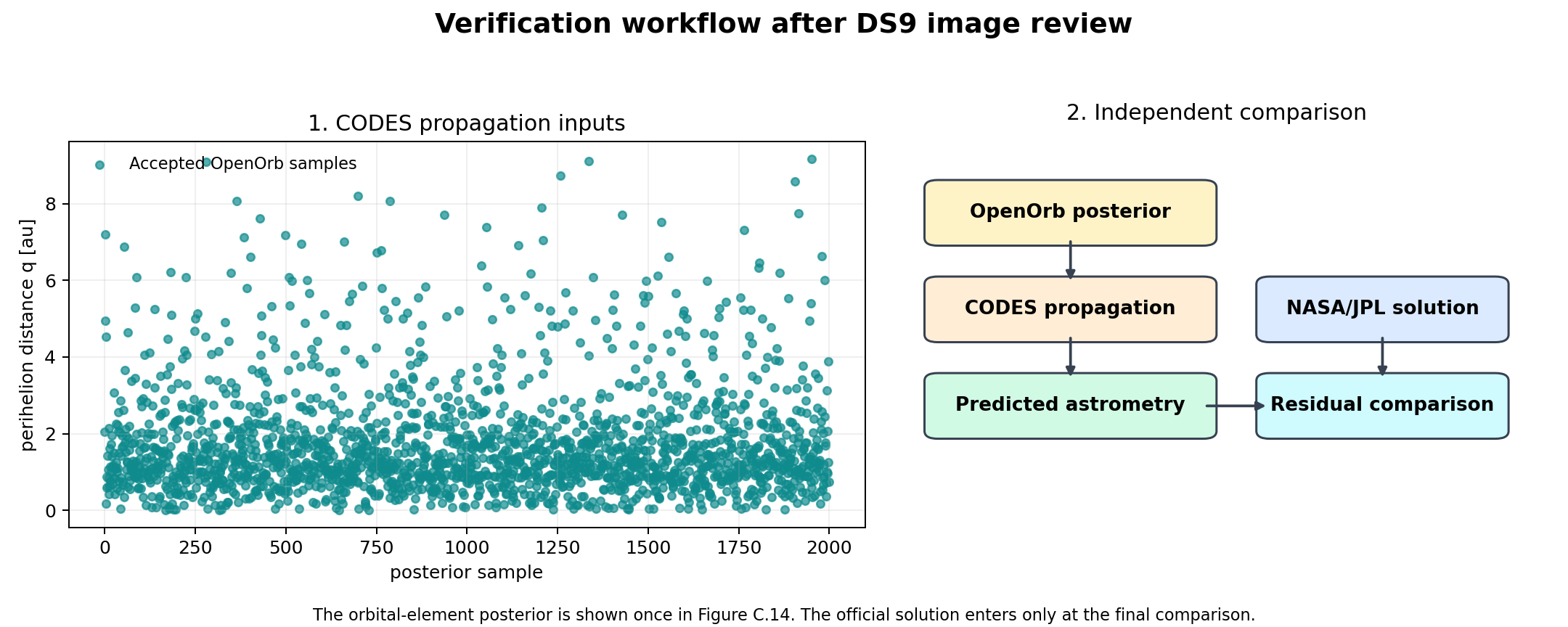}
\caption{Verification workflow after the DS9 review and OpenOrb ranging
shown in Figures~\ref{fig:ds9-ogfinder-interface} and
\ref{fig:codes-openorb-calculation}. The left panel contains only accepted
posterior samples passed to CODES. The right panel keeps the NASA/JPL
solution outside the propagation path and introduces it only for residual
comparison. A blind orbit accuracy result requires a longer arc and an
archived pre-discovery solution.}
\label{fig:ds9-codes-verification}
\end{figure}

Figures~\ref{fig:ds9-ogfinder-interface} and
\ref{fig:codes-openorb-calculation} establish the inputs to the verification.
The first figure checks the image positions. The second gives the statistical
ranging posterior. Figure~\ref{fig:ds9-codes-verification}
therefore begins with the accepted samples passed to CODES. CODES propagates
those samples with the selected planetary ephemeris and force switches.
The predicted astrometry is then compared with a separately retrieved
official solution. The official solution is not an input to propagation.
The present demonstration verifies the transfer of astrometric records,
format conversion, OpenOrb execution, preservation of the output, and CODES
ingestion. The demonstration does not claim an independent orbit determination
from five exposures.

An orbit accuracy test requires a longer observing arc, a blind fit that does
not ingest post-discovery observations, the observatory state, weighted
astrometric uncertainties, outlier rejection, and an archived reference orbit
at the discovery epoch. The requirements are consistent with the
validation boundary already stated in Section~\ref{sec:historical-ca-tests}.

\subsection{Integrated CODES desktop interface}
\label{sec:codes-gui}

The command \code{python -m neo\_orbit\_calculator} launches the desktop
interface. The interface makes the same command-line calculations available
while preserving the force switches and numerical defaults. Each action writes
the complete command to a common run log and
executes the calculation in a background thread. A long Horizons request or
Fortran propagation therefore does not freeze the window. The output
directory remains explicit for every science mode.

\begin{figure}[H]
\centering
\includegraphics[width=\textwidth]{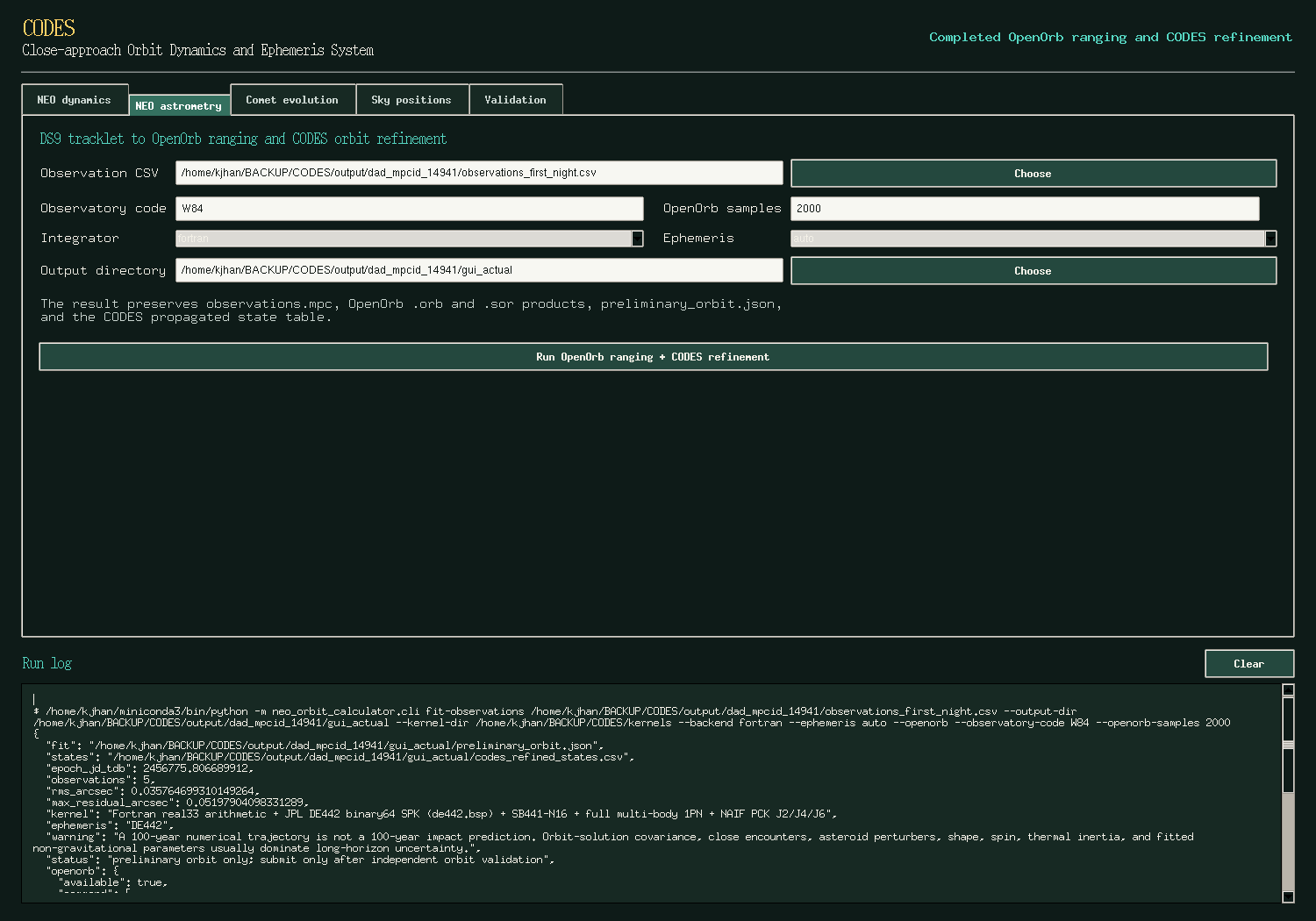}
\caption{Actual CODES desktop run of the NEO astrometry workflow. The interface
ingests five DAD/DECam positions exported by DS9 for observatory W84 and
requests 2,000 OpenOrb ranging samples. The completed run reports a
0.036\,arcsec preliminary-fit RMS and writes the OpenOrb orbit samples and
the DE442 CODES refinement. The panel is an X11 screenshot of the running Tk
application.}
\label{fig:codes-gui}
\end{figure}

\begin{table}
\caption{Operational summary of the NEO/comet dynamics GUI tabs, the
action each tab performs, and the resulting scientific data product.}
\label{tab:c6-gui-operations}
\centering
\begin{tabular}{@{}P{32mm}P{58mm}P{64mm}@{}}
\toprule
\color{cGold}\textbf{GUI tab} &
\color{cGold}\textbf{Operation} &
\color{cGold}\textbf{Scientific product}\\
\midrule
\textbf{NEO dynamics} &
Selects the designation, TDB interval, sampling, covariance clone count and
seed, force terms, and local non-gravitational coefficients. Separate buttons
request an SPK, retrieve Horizons vectors, run the nominal propagator, or run
the covariance ensemble. &
Authoritative binary SPK, sampled state-vector CSV, CODES orbit projection,
full-covariance clone products, and residual summary against Horizons\\
\midrule
\textbf{NEO astrometry} &
Imports a DS9 astrometry CSV with UTC, right ascension, declination,
observatory code, ranging sample count, integrator, and ephemeris. The action
runs OpenOrb statistical ranging before the CODES refinement. &
MPC-format observations, OpenOrb \code{.sor} and \code{.orb} samples,
preliminary orbit JSON, CODES refined-state CSV, and logged residual
statistics\\
\midrule
\textbf{Comet evolution} &
Selects a comet and an apparition-year interval. Known return years may be
entered when Horizons aliases are incomplete. &
CSV of the separate JPL apparition solutions and a plot of return interval,
semimajor axis, and perihelion distance\\
\midrule
\textbf{Sky positions} &
Accepts UTC epochs and a Horizons observer code for a selected comet. &
Apparent right ascension, declination, heliocentric distance, observer
distance, IAU constellation, and a labeled sky track\\
\midrule
\textbf{Validation} &
Runs the ten-object NEO close-approach comparison or the historical
1P/Halley return comparison. &
Traceable CSV and JSON results, diagnostic figures, and direct residuals
against NASA/JPL or observation-constrained historical solutions\\
\bottomrule
\end{tabular}
\end{table}

The GUI does not convert a sensitivity integration into an operational hazard
solution. Impact probability still requires a fitted covariance and virtual
asteroid propagation. Separate commands preserve the provenance of
authoritative JPL retrieval, CODES propagation, and validation.

\section{Five-Year Reference Allocation}
\label{sec:neo-five-year-schedule}

The shared mission scheduler assigns the Volume IV program within the same
260-week capacity model used for the other science appendices. The current
reference scenario reserves 2184.0 hr for an optical blind survey and 873.6 hr
for alert-driven recovery. The 3057.6 hr total equals 6.98 per cent of the
five-year wall clock. Table~\ref{tab:neo-five-year} separates the two purposes.

\begin{table}[H]
\centering
\caption{Database-driven five-year reference allocation for Volume IV. The
blind-survey line is a capacity scenario rather than an adopted completeness
claim.}
\label{tab:neo-five-year}
\begin{tabularx}{\textwidth}{@{}P{43mm}C{25mm}C{25mm}Y@{}}
\toprule
\color{cGold}\textbf{Program} &
\color{cGold}\textbf{Mission years} &
\color{cGold}\textbf{Time [hr]} &
\color{cGold}\textbf{Interpretation}\\
\midrule
NEO optical blind survey & 1--5 & 2184.0 & Five per cent of the mission wall
clock used to test available survey capacity outside commissioning and
Director weeks\\
NEO recovery ToO reserve & 1--5 & 873.6 & Two per cent of the mission wall
clock retained for externally discovered objects with measured astrometry\\
\midrule
\textbf{Volume IV total} & \textbf{1--5} & \textbf{3057.6} &
\textbf{6.98 per cent of the five-year wall clock}\\
\bottomrule
\end{tabularx}
\end{table}

\begin{figure}[H]
\centering
\includegraphics[width=\textwidth]{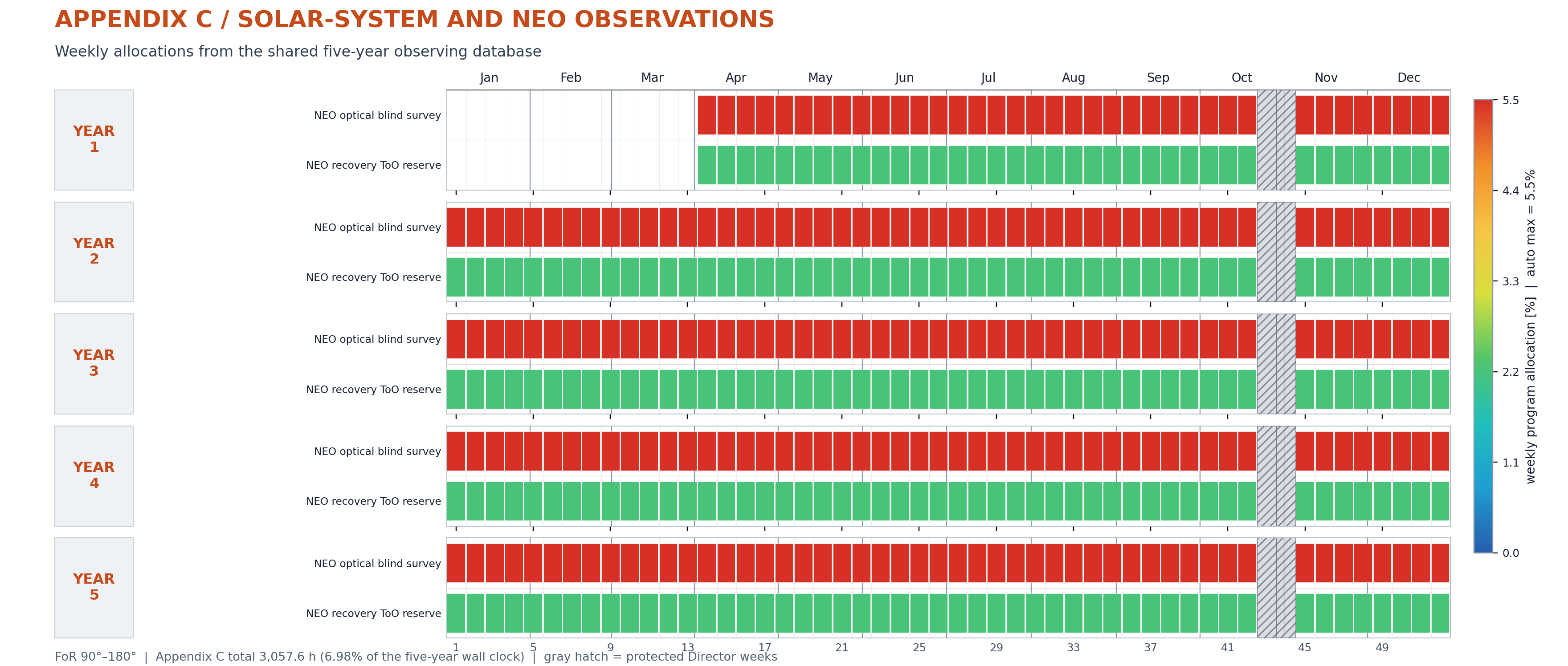}
\caption{Weekly Volume IV allocation generated from the shared observing
database. Color gives the fraction of one 168-hr week assigned to each
program. The color scale uses the largest Volume IV allocation and gray
hatching marks the protected Director weeks. The uniform recovery row
represents reserved capacity rather than a preselected target list.}
\label{fig:neo-five-year}
\end{figure}

Figure~\ref{fig:neo-five-year} shows the weekly allocation. Routine science
begins in Year~1 week~14 after commissioning and performance acceptance. The
scheduler excludes Director weeks 43 and 44 in every year and applies a
$90^\circ$--$180^\circ$ solar-elongation field of regard. The blind-survey row
demonstrates that the shared schedule can reserve the requested time. The
allocation does not demonstrate a competitive NEO census or the completeness of a fixed
survey footprint. The recovery row remains unassigned until an alert supplies
an ephemeris and covariance.

\section{Mission Decision Matrix}

The decision matrix separates measurements supported by the baseline payload
from measurements that require a wavelength extension or a ground-based
partner. The comparison also identifies the dominant limitation of each
configuration.

\begin{table}
\caption{Scientific strength and principal limitation of each NEO
characterization configuration, from the 0.2--1.5\,$\mu$m baseline
through ground-based MIR and external-survey coordination.}
\label{tab:c11-decision-matrix}
\centering
\begin{tabularx}{\textwidth}{@{}P{35mm}P{47mm}Y@{}}
\toprule
\color{cGold}\textbf{Configuration} &
\color{cGold}\textbf{Scientific strength} &
\color{cGold}\textbf{Principal limitation}\\
\midrule
Baseline 0.2--1.5\,$\mu$m &
Rapid astrometry, recovery, light curves, phase curves, and partial
visible/NIR taxonomy &
Diameter--albedo degeneracy remains. Discovery close to the Sun is limited
by field of regard.\\
Extended to 2.45\,$\mu$m &
More complete Bus--DeMeo coverage and improved mineralogical continuity &
Still dominated by reflected light for ordinary NEO temperatures\\
Ground-based MIR coordination &
Thermal flux provides diameter and albedo constraints for selected targets &
Atmospheric windows, thermal background, weather, and target ephemeris limit
the available observations\\
External-survey follow-up &
High scientific return per unit mission complexity through coordination with
Rubin, radar, ground-based astrometry, and the MPC &
Discovery selection function is inherited from external facilities\\
\bottomrule
\end{tabularx}
\end{table}

\section{Recommended Proposal Position}

The proposal should claim the measurements supported by the adopted payload
and reserve population-yield claims for a validated survey simulation. The
following commitments define that boundary.

\begin{enumerate}[leftmargin=7mm,itemsep=2.2mm]
\item Commit the baseline mission to rapid NEO follow-up, ICRF astrometry,
time-resolved photometry, and visible/NIR characterization.
\item Exclude a 4--10\,$\mu$m detector, MIR optics, MIR ETC predictions, and
MIR discovery yields from the 3.5ST payload and performance claims.
\item Establish ground-based MIR coordination for selected objects. The
agreement must define response time, atmospheric windows, flux calibration,
filter transmission, timing metadata, and delivery of covariance information.
\item Use at least four astrometric measurements in one recovery visit.
Select the exposure time and repeat interval from angular rate and orbit
covariance.
\item Do not claim a competitive blind NEO census with the baseline field.
Quantify the characterized-target yield by applying the 3.5ST
Target-of-Opportunity allocation and ground MIR availability to an external
alert stream.
\item Preserve a $60^\circ$ baseline solar avoidance requirement until a
thermal and stray-light model demonstrates access to $45^\circ$.
\item Validate orbit products against Horizons SPK output. Use covariance
propagation before making any impact-probability claim.
\item Use $GM$ values and versioned JPL kernels throughout. Record kernel
checksums and orbit-solution identifiers in every derived product.
\end{enumerate}

\begin{warningbox}
\textbf{\color{cRed}Claim boundary.}
The optical single-visit calculation establishes a reference flux limit and
an idealized warning geometry. The calculation does not establish blind
survey completeness. A defensible program yield requires a measured optical
and near-infrared response, the final field of regard, a deterministic
Target-of-Opportunity allocation, synthetic-source recovery, false-link
control, orbit-quality cuts, and the availability of ground MIR observations.
\end{warningbox}

\clearpage
\ifdefined\APPENDIXCBOOK
\renewcommand{\bibname}{References}
\else
\renewcommand{\refname}{References}
\fi

\appendixCfinish

\let\APPENDIXCBOOK\undefined

\clearpage
\definecolor{Blue1}{HTML}{1FABD5}
\definecolor{Blue2}{HTML}{1D8DB0}
\definecolor{Blue3}{HTML}{116E8A}
\backmatter
\MakeBackCover
\end{document}